\documentclass[12pt]{article}
\newcommand{\blind}{0}

\RequirePackage{amsthm,amsmath,amsfonts,amssymb}
\RequirePackage[authoryear]{natbib}
\RequirePackage[colorlinks,citecolor=blue,urlcolor=blue]{hyperref}%% uncomment this for coloring bibliography citations and linked URLs
\RequirePackage{graphicx}%% uncomment this for including figures
\usepackage{booktabs, algorithmic, rotating, enumitem, adjustbox, bm} 
\usepackage{array,stmaryrd}
\usepackage{tabularx,comment}
\usepackage{pdflscape}
\usepackage{adjustbox}
\usepackage{subfig}
\usepackage{caption}
\usepackage{longtable}
\usepackage{listings}
\usepackage{multirow}
\usepackage{xr}
\usepackage{blkarray}
\usepackage{xcolor}
\usepackage{setspace}% http://ctan.org/pkg/setspace
\usepackage[margin=1in]{geometry}

\newcommand{\E}{\mathop{\mathbb{E}}} % expectation
\begin{document}
\def\spacingset#1{\renewcommand{\baselinestretch}%
{#1}\small\normalsize} \spacingset{1}

\if0\blind
{
  \title{\bf Generalized Multilevel Functional Principal Component Analysis with Application to  NHANES Active Inactive Patterns}
  \author{Xinkai Zhou\thanks{
    This work was supported by the National Institute of Neurological Disorders and Stroke under Award Number R01 NS060910, and National Institute on Aging under Award Number R01 AG075883.}\hspace{.2cm}\\
    \small{Department of Biostatistics, Johns Hopkins University}\\
    Julia Wrobel \\
    \small{Department of Biostatistics and Bionformatics, Emory University}\\
    Ciprian M. Crainiceanu \\
    \small{Department of Biostatistics, Johns Hopkins University}\\
    Andrew Leroux\\
    \small{Department of Biostatistics and Informatics, Colorado School of Public Health}}
    \date{}
  \maketitle
} \fi

\if1\blind
{
  \bigskip
  \bigskip
  \bigskip
  \begin{center}
    {\LARGE\bf Analysis of Active/Inactive Patterns in the NHANES Data using Generalized Multilevel Functional Principal Component Analysis}
\end{center}
  \medskip
} \fi

\bigskip

\begin{abstract}
Between 2011 and 2014  NHANES collected objectively measured physical activity data using wrist-worn accelerometers for tens of thousands of individuals for up to seven days. In this study, we analyze minute-level indicators of being active, which can be viewed as binary (since each minute is either active or inactive), multilevel (because there are multiple days of data for each participant), and functional data (because the within-day measurements can be viewed as a function of time). To identify both within- and between-participant directions of variation in these data, we introduce Generalized Multilevel Functional Principal Component Analysis (GM-FPCA), an approach based on the dimension reduction of the linear predictor.  Our results indicate that specific activity patterns captured by GM-FPCA are strongly associated with mortality risk. Extensive simulation studies demonstrate that GM-FPCA accurately estimates model parameters, is computationally stable, and scales up with the number of study participants, visits, and observations per visit. \texttt{R} code for implementing the method is provided.
\end{abstract}

\noindent%
\vfill

\newpage
\spacingset{1} 
\section{Introduction}
Body-worn accelerometers enable high-resolution, objective measurement of human physical activity. For example, both the US National Health and Nutrition Examination Survey (NHANES) and the UK Biobank study collected accelerometry data for tens of thousands of study participants.  This paper focuses on the NHANES 2011-2014 accelerometry data, publicly released in December 2020, which contain minute level summaries of physical activity for $13,603$ participants over seven days collected from wrist-worn accelerometers. Physical activity at each minute is summarized using a continuous measure called Monitor Independent Movement Summary units (MIMS) \citep{john2019mims}. Because MIMS can be difficult to interpret directly, they are often transformed to improve translation (e.g., it makes sense to recommend decreasing the number of minutes sedentary, while no one knows how to increase their daily activity by $100$ MIMS). Possible transformations include (1) binary data corresponding to active/inactive, walking/non-walking, or sleep/wake; (2) multinomial data distinguishing sedentary, low-intensity physical activity (LIPA), and moderate-to-vigorous physical activity (MVPA); and (3) count data capturing the number of steps or active seconds per minute. This paper focuses on the binary active/inactive profile derived by thresholding MIMS, but our method applies more generally to other non-Gaussian data, including discrete data (e.g., binary, multinomial, counts) \citep{gaston2008,Kass2001,Kelly2012,sebastian2010,senturk2014,swihart2015, bothwell2022pattern,cui2022fui} or continuous data with strong departures from normality \citep{gaynanova2020,staicu2012}. 
 
\begin{figure}[!tbp]
\centering
\includegraphics[width=\textwidth]{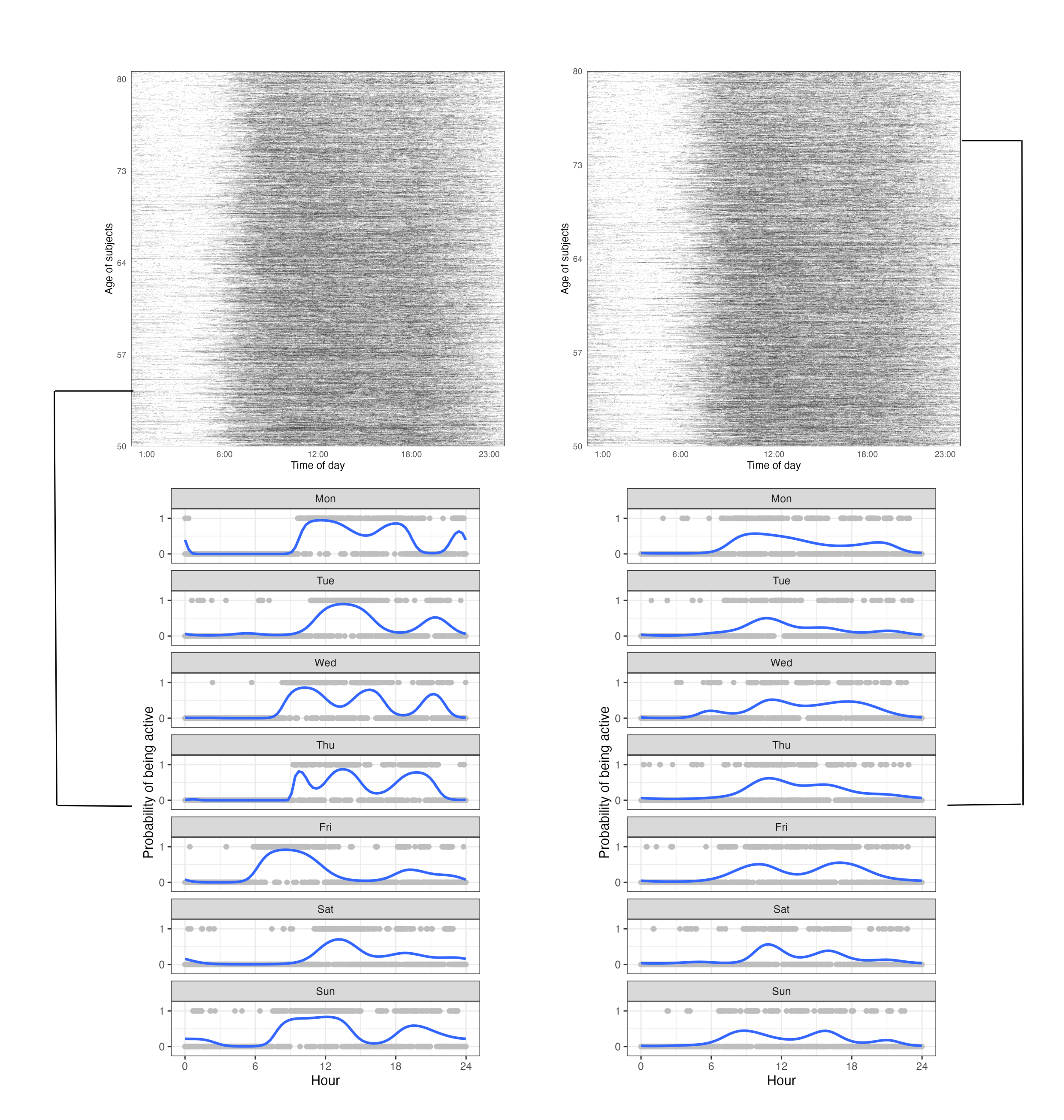}
  \caption{Top panels: binary active (black dots) / inactive (white dots) profiles for $4,445$ NHANES participants aged 50 years or older, shown for Wednesday (left) and Saturday (right). . Bottom panels: two randomly selected participants; gray dots are minute-level data, and blue curves are smoothed probabilities of being active.}
  \label{fig:nhanes-allsubjects}
\end{figure}

Figure~\ref{fig:nhanes-allsubjects} displays the binary physical activity data for the $4,445$ subjects who participated in the NHANES 2011-2014 accelerometry study and are 50 years or older.  The binary active/inactive data is obtained by thresholding the MIMS data as $Z_{ijk} = \mathbf{1}_{\{W_{ijk} \geq 10.558\}}$, where $W_{ijk}$ corresponds to the $i^{\text{th}}$ individual's MIMS measurement on day $j=1,\ldots,J_i$ at minute $k$. The threshold of $10.558$ for active versus inactive was suggested by \citet{karas2022comparison}. The top panel shows the active minutes (black) versus inactive (white) from midnight to midnight for Wednesday and Saturday. The bottom panels show day-to-day variation within two randomly selected participants. Each sub-panel corresponds to one day and displays both the binary data (gray dots) and smooth estimates of the probability of being active (blue curves) \citep{wood2017gam}. These data illustrates the large heterogeneity between- and within-subjects, though some population level structure is apparent (e.g., less activity at night), while some may be hidden. Our goal is to model such generalized (e.g., binary) functional (e.g., minutes within a day) data observed at multiple time points (e.g., days).

We are not aware of any existing methods that can handle data of the size and complexity illustrated in Figure~\ref{fig:nhanes-allsubjects}. Indeed, the binary active/inactive profiles in the NHANES data exhibit the following characteristics:  (1) a large number of study participants ($4,445$) and observations per day ($1,440$) (2) multiple days of observation (from $1$ to $7$); (3) substantial within- and between-day variability;  (4) high within-participant correlation over the course of the day; and (5) binary observations. To address the combination of these challenges we introduce generalized multilevel functional principal component analysis (GM-FPCA). Here ``generalized" refers to the non-Gaussianity of the measurements, ``multilevel" refers to repeated visits within participants, and ``functional" refers to the representation of data as curves over time. In our case,  each day of physical activity trajectory from  each participant is treated as a separate functional observation. 

The paper is organized as follows. Section~\ref{sec:literature-review} reviews related methods for functional data. Section~\ref{sec:methods} presents GM-FPCA. Section~\ref{sec:simulation} uses simulation to assess statistical and computational performance. Section~\ref{sec:real-data} applies GM-FPCA to the NHANES 2011--2014 accelerometry data. Finally, Section~\ref{sec:discussion} provides concluding remarks.

\section{Literature Review}\label{sec:literature-review}
Methodological and computational advances in functional data analysis have allowed the extension of method to data sets that are increasingly more complex. Our own approach, GM-FPCA, builds upon three methods in functional data analysis: (1) functional principal components analysis (FPCA); (2) generalized functional principal component analysis (GFPCA) for generalized single-level functional data; and (3)  multilevel functional principal component analysis (MFPCA) for Gaussian functional data repeatedly measured for the same study participant (e.g., minute level physical activity data measured over multiple days). While a vast literature and reliable software exist for FPCA \citep{castro1986principal, Staniswalis1998,jones1992displaying,ramsaysilv2005,morris2015,xiao2016,refund,crainiceanu2023book}, Generalized (non-Gaussian measurements) and Multilevel (repeatedly measured functional outcomes) FPCA require a closer assessment of the current state-of-the-art. 

Generalized functional principal components analysis (GFPCA) extends FPCA to exponential-family outcomes such as binary or count data. More precisely, let $Z_{ik}$ denote the non-Gaussian functional observation for participant $i = 1, \ldots, I$ at a point $s_k\in \mathcal{S}$, where $\mathcal{S}$ is a compact domain. The GFPCA model is 
\begin{equation}
g\{\E(Z_{ik})\}= \mu_0(s_k) + a_i(s_k) = \mu_0(s_k) + \sum_{l=1}^\infty \xi_{il}\phi_l(s_k)\;,
    \label{eqn:gfpca}
\end{equation}
where $g(\cdot)$ is a link function, $\mu_0(s)$ is the population-mean function on the linear predictor scale, $a_i(s)$ is a zero-mean Gaussian process which, by the Kosambi–Karhunen–Lo\`eve theorem, can be expanded as $a_i(s) = \sum_{l=1}^\infty \xi_{il}\phi_l(s)$, where $\{\phi_l(s)\}_l$ are population-level orthonormal eigenfunctions satisfying $\int_S \phi_l^2(s) ds = 1$, and the scores $\xi_{il} \sim N(0, \lambda_l)$ are subject-specific and uncorrelated.

Several estimation methods have been proposed for Model (\ref{eqn:gfpca}). Early work by \citet{hall2008gfpca} modeled $Z_{ik}$ via a latent Gaussian process with mean and covariance estimators constructed using the Taylor expansion of $g(\cdot)$. Bayesian approaches for non-Gaussian functional data were introduced by \citet{Gertheiss2017gfpca} and \citet{linde2009gfpca} under various sampling scenarios. \citet{li2018exponential} proposed Exponential Family Functional Principal Component Analysis (EFPCA), employing quadratic penalties on the scores and eigenfunctions of a low-rank decomposition of the linear predictor. \citet{Weishampel2023} proposed first smoothing non-Gaussian data at the subject level and then use  FPCA on the estimated trajectories on the linear predictor scale. Here we will focus on tested methods that have been implemented in reproducible software: (1) a variational Bayes approach for binary functional data \citep{wrobel2019registration}, implemented in R via the function {\ttfamily registr::bfpca()}  \citep{registr_package,registr2}; (2) the two-step GFPCA method described in \citep{Gertheiss2017gfpca}, implemented in the function {\ttfamily registr::gfpca\_twoStep()}; and (3) a fast GFPCA method \citep{leroux2023fast}, deployed in the {\ttfamily fastGFPCA} package \citep{fastGFPCA_software}. The fast GFPCA approach estimates the latent processes in small bins along the functional domain, which allows to borrow strength across study participants irrespective of the type of outcome. The  eigenfunctions $\{\phi_l(s)\}$ are then estimated by smoothing these initial estimators using the fast covariance estimation (FACE) algorithm \citep{xiao2016face}, while the scores $\{\xi_{il}\}$ are estimated using a global generalized linear mixed model (GLMM) conditional on the  smooth estimators of the eigenfunctions $\{\phi_l(s)\}$. 

Multilevel functional principal components analysis (MFPCA) extends FPCA to the case when  Gaussian functional data are observed at multiple visits (e.g., days). MFPCA was first introduced by \citet{Di2009mfpca} and its implementation is available in the function {\ttfamily refund::mfpca.sc()}. A faster version was recently released in {\ttfamily refund::mfpca.face()}, based on the work of \citet{cui2022mfpca}. Further developments include extensions to multivariate and multilevel functional data \citep{Zhang2023MMFPCA}.

Extending these models to accommodate non-Gaussian outcomes and a multilevel functional structure simultaneously is not straightforward, though several strategies have been proposed. 
For instance, \citet{chen2013gmfpca} described a reduced-rank method for generalized functional data. They used a marginal approach that does not estimate the random effects, which are the major  focus of our framework. A series of papers and software \citep{ivanescu2015,scheipl2015functional,Scheipl2016GFADM,refund} introduced the current state-of-the-art  for generalized function-on-scalar regression, which is designed to handle multilevel subject-specific random effects. This approach can be slow for a larger number of study participants and, in our experience, does not yet scale up to our NHANES application. This may change in the future, and having multiple approaches based on different inferential structures can be helpful in complex data analytic scenarios. \citet{goldsmith2015gmfpca} proposed a generalized multilevel function-on-scalar regression using FPCA on the underlying latent space.  When applied to a physical activity dataset of $583$ subjects with five days per subject and $144$ observations per day (one observation for every 10 minutes), the method took 10 days to run, so we could not use this method on our NHANES data, which contains thousands of participants, multiple days per participant, and $1,440$ observations per day (one observation per minute). 

Therefore, both new methods and associated  computational tools need to be developed both to account for the complexity of the NHANES data set, but also to prepare for even larger and more complex data sets. Below we describe our method and show how it can be effectively implemented.

\section{Methods}\label{sec:methods}
The observed data are of the form $\{s_k, Z_{ijk}\}$ and $Z_{ijk}$ is the non-Gaussian observation for subject $i = 1, ..., I$, at visit $j = 1, ..., J$, and domain point $s_k\in\mathcal{S}$.  While the data are  observed on a discrete grid, we assume that they are generated by the GM-FPCA model 
\begin{eqnarray}
   g[\E\{Z_{ijk}\}] = \mu_0(s_k) + \mu_j(s_k) + a_i(s_k) + b_{ij}(s_k)\;. 
   \label{eqn:global-model}
\end{eqnarray}
Here $g(\cdot)$ is a link function, $\mu_0(\cdot)$ is the population-mean function on the linear predictor scale, $\mu_j(\cdot)$ is the visit-specific shift from the population mean, and $a_i(\cdot)$ and $b_{ij}(\cdot)$ are subject- and subject-visit-specific deviations from the population-mean function. We further assume that $a_i(s) \sim GP(0, K_a)$ and $ b_{ij}(s) \sim GP(0, K_b)$ 
are mutually independent zero-mean Gaussian processes with covariance operators $K_a$ and $K_b$, respectively. 

The main goal of GM-FPCA is to decompose the variability in $a_i(s)$ and $b_{ij}(s)$ along their main directions of variation. By the Kosambi–Karhunen–Lo\`eve theorem, 
\begin{eqnarray*}
    a_i(s) = \sum_{l=1}^\infty \xi_{il}\phi_l(s)\;, \,\,
    b_{ij}(s) = \sum_{m=1}^\infty \zeta_{ijm}\psi_m(s)\;,
\end{eqnarray*}
where $\{\phi_l(\cdot)\}$ and $\xi_{il} \sim N(0, \lambda_l^{(1)})$, $l = 1, 2, \ldots$  are subject-level orthonormal eigenfunctions and scores, $\{\psi_m(\cdot)\}$ and $\zeta_{ijm} \sim N(0, \lambda_m^{(2)})$, $m = 1, 2, \dots$ are subject-visit level orthonormal eigenfunctions and scores, and the scores $\xi_{il}$ and $\zeta_{ijm}$ are mutually independent. With these assumptions the GM-FPCA model becomes
\begin{equation}
     g\{E(Z_{ijk})\} = \mu_0(s_k) + \mu_j(s_k) + \sum_{l=1}^\infty \xi_{il}\phi_l(s_k) + \sum_{m=1}^\infty \zeta_{ijm}\psi_m(s_k)\;.
    \label{eqn:gmfpca}
\end{equation}

Our approach shares similarities with local polynomial  regression models \citep{fan1996local}. Specifically, we use a sliding ``window" across the domain, forming data bins that contain the local information from all study participants. In each bin, we fit a \emph{local} generalized linear mixed model (GLMM) with both subject- and subject-visit random intercepts; they are then used as local approximations to the true subject- and subject-visit random deviation functions $a_i(s_k)$ and $b_{ij}(s_k)$. The local estimators of $a_i(s_k)+b_{ij}(s_k)$ are then decomposed using MFPCA to estimate the subject- and subject-visit level eigenfunctions $\phi_l(\cdot)$ and $\psi_m(\cdot)$ over the entire domain. Finally, we use a Bayesian multilevel model to estimate the scores $\xi_{il}$ and $\zeta_{ijm}$ on the full data without binning and conditional on the eigenfunction estimators. Below we provided the  details for each step.

\emph{Step 1: Data Binning}. We partition the domain $\mathcal{S}$ into overlapping bins. A sliding window of width $w$ moves across $\mathcal{S}$, forming bins centered at each $s_k$. Denote these bins by $S_k$, where each bin $S_k$ contains domain points $\{s_{k-\lceil \frac w2 \rceil}, ..., s_k, ..., s_{k+\lceil \frac w2 \rceil}\}$. On the domain boundary, if the data are cyclic (e.g., 24-hour activity data), bins can wrap around. Otherwise, boundary bins contain fewer points. In addition to partitioning the problem into smaller, manageable subproblems, binning offers key advantages when combined with Step 2, as explained below. 
frameframeframe
\emph{Step 2: Local GLMM Fitting}. Within each bin $S_k$, we fit a GLMM of the form
\begin{eqnarray}\label{eqn:local-model}
g[\E\{Z_{ijk}\mid k: s_k\in S_k\}] = \mu_{0k}^* + \mu_{jk}^* + a_{ik}^* + b_{ijk}^*
\end{eqnarray}
where $\mu_{0k}^*$ and $\mu_{jk}^*$ are fixed effects, and $a_{ik}^*$ and $b_{ijk}^*$ are subject- and subject-visit-specific random intercepts, respectively. Estimates of these local quantities (denoted by the $^*$) approximate their functional counterparts defined in Model \ref{eqn:global-model}. Specifically, functional random effects at the original sampling points $s_1, ..., s_K$ are approximated as $a_i(s_k) = \widehat{a}_{ik}^*$ and $b_{ij}(s_k) = \widehat{b}_{ijk}^*$. We extract the linear predictor $\widehat{\eta}_{ijk}^* = \widehat{\mu}_{0k}^* + \widehat{\mu}_{jk}^* + \widehat{a}_{ik}^* + \widehat{b}_{ijk}^*$ for decomposition in Step 3.

Without binning, the subject-visit-specific random intercept $b_{ijk}^*$ would not be estimable, because there is only one observation for each subject-visit pair at each domain point. This is a key difference between our approach and the pointwise fitting approach introduced by  \citet{cui2022fui}. Furthermore, binning improves estimation stability in potentially sparse or imbalanced settings (e.g., class imbalance in binary outcomes). Fitting local GLMMs also leverages the ability of GLMMs to handle missing data and borrow information from neighboring observations as well as across subjects and visits.

\emph{Step 3: Decomposition of local estimates using MFPCA}. In this step we start with the estimated linear predictors from local models, $\widehat{\eta}_{ijk}^*$,  and apply MFPCA \citep{Di2009mfpca} to decompose them into subject- and subject-visit-specific components along the functional domain. We use the fast MFPCA approach \citep{cui2022mfpca} to obtain smooth estimates of the eigenfunctions, which we denote by $\{\widehat{\phi}_l(s)\}_{l=1}^L$ and $\{\widehat{\psi}_m(s)\}_{m=1}^M$. The choice of $L$ and $M$ can be guided by criteria such as explained variance and is further discussed in the supplementary material.

\emph{Step 4: Global score estimation via Bayesian multilevel modeling}. Conditional on the eigenfunctions $\{\widehat{\phi}_l(s)\}$ and $\{\widehat{\psi}_m(s)\}$, the scores $\{\xi_{il}\}$ and $\{\zeta_{ijm}\}$ are estimated using  the global GLMM:
\begin{eqnarray}\label{eqn:global-glmm}
g[\E\{Z_{ijk}\}] = \mu_0(s_k) + \mu_j(s_k) + \sum_{l=1}^L \xi_{il}\widehat \phi_l(s_k) + \sum_{m=1}^M \zeta_{ijm}\widehat \psi_m(s_k).
\end{eqnarray}
which includes random slopes for the eigenfunctions. This model has important characteristics that provide substantial computational advantages: (1) both the number of subject-specific ($L$) and visit-specific ($M$) random effects is relatively small, controlling the complexity of the model; (2) the eigenfunctions are orthonormal at each level, which improves numerical stability and simplification of algorithms; and (3) the global model is a GLMM, which provides all the benefits of the GLMM inferential machinery, including theoretical properties.

The global model~\eqref{eqn:global-glmm} is the simplest possible model that accounts for the non-Gaussian, functional, and multilevel structure of the data. However, fitting it using standard software (e.g., \verb|lme4| \citep{bateslme4}) can be computationally prohibitive for even moderately large data sizes (e.g., $1000$ subjects, $10$ visits per subject, and $100$ sampling points). This likely explains why more complex models designed for the same problem cannot currently fit data sets of the size and complexity considered here. Therefore, we adopt a Bayesian multilevel approach, which takes advantage of the simple structure of the full conditional distributions implied by model~\eqref{eqn:global-glmm}.  For example, the full conditional for the random effect $[\zeta_{ijm}\mid{\rm others}]$ is proportional to $$[Z_{ijk}\mid\zeta_{ijm},{\rm others}][\zeta_{ijm}, {\rm others}]\;,$$ which does not have a closed-form, but are relatively easy to sample using modern Bayesian software such as \verb|JAGS| \citep{hornik2003jags} or \verb|Stan| \citep{carpenter2017stan}. 

In the supplementary materials, we provide the \texttt{R} code that implements the method and discuss practical considerations such as selecting the bin width and determining the number of eigenfunctions. These choices generally depend on data size, sampling density, and desired computational speed. Currently we recommend a sensitivity analysis to these choices.

\section{Simulation Study}\label{sec:simulation}
In this section, we simulated binary and Poisson multilevel functional data to evaluate how well the proposed method recovers the eigenfunctions, scores, and latent functional trajectories. For both binary and Poisson data, the latent continuous response $\eta_{ijk}$ was generated from the following model:
\begin{eqnarray*}
\eta_{ijk} &=& \mu_0(s_k)+ \sum_{l=1}^4 \xi_{il}\phi_l(s_k) + \sum_{m=1}^4 \zeta_{ijm}\psi_m(s_k),
\end{eqnarray*}
where $\mu_0(s_k)$ denotes the intercept at time $s_k$, $\xi_{il} \sim N(0, \lambda_l^{(1)})$, $\zeta_{ijm}\sim N(0, \lambda_m^{(2})$,  $\{s_k = \frac{k}{K}: k = 0, 1, ..., K\}$, and $K$ is the number of sampling points. We set the true eigenvalues $\lambda_l^{(1)} = 0.5^{l-1}, l = 1, 2, 3, 4$ and $\lambda_m^{(2)} = 0.5^{m-1}, m = 1, 2, 3, 4$. Following \citet{Di2009mfpca}, we considered two cases for simulating eigenfunctions:
\newline

\noindent\textit{Case 1.} Mutually orthogonal bases.
\begin{enumerate}[label=Level \arabic*:, align=left, leftmargin=0.5in]
    \item [Level 1:] $\phi_l(s) = \{\sqrt{2} \sin(2\pi s), \sqrt{2} \cos(2\pi s), \sqrt{2} \sin(4\pi s), \sqrt{2} \cos(4\pi s)\}.$
    \item [Level 2:] $\psi_m(s) = \{\sqrt{2} \sin(6\pi s), \sqrt{2} \cos(6\pi s), \sqrt{2} \sin(8\pi s), \sqrt{2} \cos(8\pi s)\}.$
\end{enumerate}
\noindent\textit{Case 2.} Mutually nonorthogonal bases between the two levels.
\begin{enumerate}[label=Level \arabic*:, align=left, leftmargin=0.5in]
    \item same as Case 1.
    \item $\phi_1^{(2)}(s) = 1, \phi_2^{(2)}(s) = \sqrt 3 (2s-1), \phi_3^{(2)}(s) = \sqrt 5 (6s^2-6s+1), \phi_4^{(2)}(s) = \sqrt 7 (20s^3 - 30s^2 + 12s - 1)$.
\end{enumerate}

In addition to varying the eigenfunctions, we also examined the effect of sample size $I=50, 100, 200, 500, 1000$, number of visits $J=2, 5, 10$,  number of sampling points $K=100, 200, 500$, bin length corresponding to  $w=2\%, 5\%, 10\%, 15\%, 20\%$ of the number of sampling points, and fixed-effect intercept function $\mu_0(s)=b_0 = 0, -1.5, -2.5, -3.5$ for binary data, corresponding roughly to $50\%, 30\%, 17\%,$ and $9\%$ ones. Except for eigenfunctions, we varied one parameter at a time and held others fixed to avoid an infeasible number of scenarios. For each scenario we conducted $20$ simulations. The number of post warm-up sampling iterations was set to $1,000$ for fitting the Bayesian model in Step 4.

To evaluate the proposed method, we calculate the mean squared error (MSE) of the linear predictors across all subject-visit pairs
$$
\frac{1}{I\times J\times K} \sum_{i=1}^I\sum_{j=1}^J \sum_{k=1}^K[\widehat\eta_{ijk} - \eta_{ijk}]^2,
$$
and the integrated squared error (ISE) of each eigenfunction: 
$$
\int_{0}^1\{\widehat\phi_{m}^{(l)}(s) - \phi_{m}^{(l)}(s)\}^2 ds,
$$
for $l = 1, 2$ (two levels) and $m = 1, 2, 3, 4$. 

\subsection{Simulation Results}
\begin{figure}[!tbh]
\centering
\includegraphics[width=0.7\textwidth]{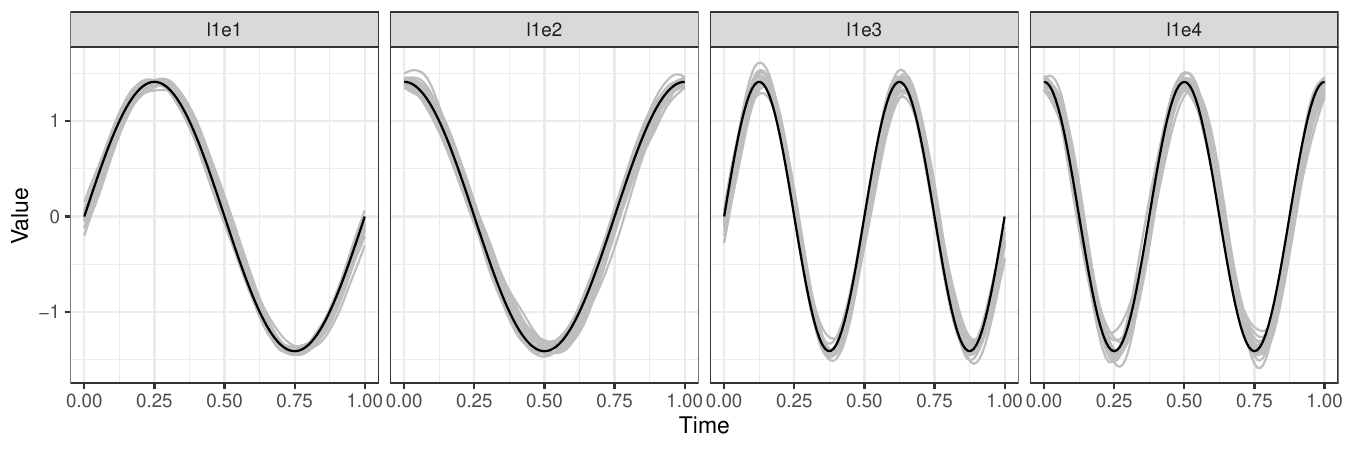}
\includegraphics[width=0.7\textwidth]{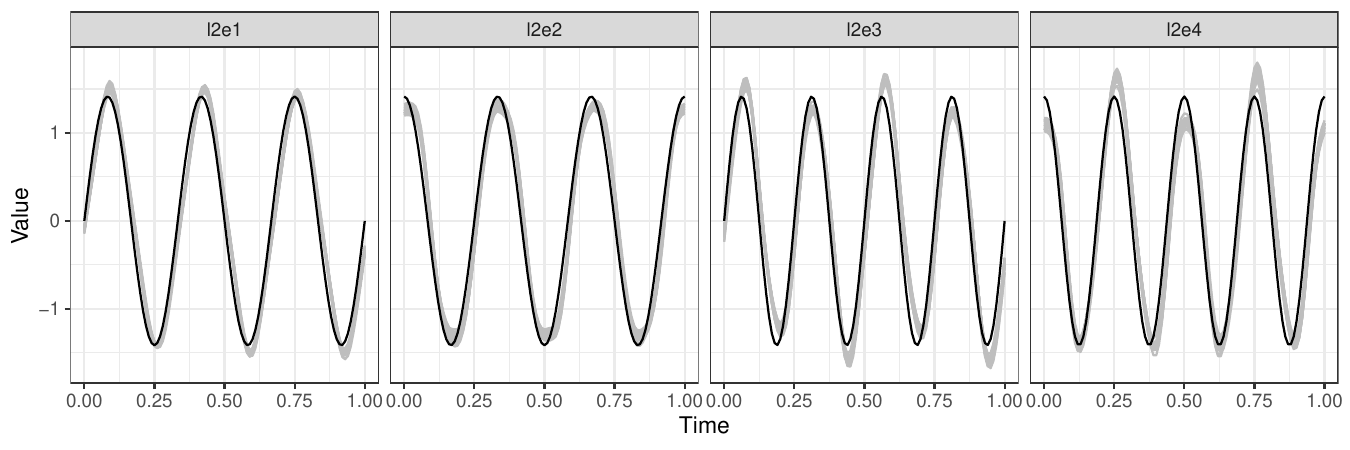}
  \caption{Estimated (gray) versus true eigenfunctions (black) from 20 simulations for the setting of binary data, Case 1 eigenfunctions, sample size $I = 1000$, number of visits $J = 10$, number of sampling points $K = 100$, bin width corresponding to $w = 5\%$ of the sampling points, and fixed effect intercept $b_0 = 0$. The panel label ``l1e1" refers to the first eigenfunction at level 1. Similarly for other labels.}
  \label{fig:efs-example}
\end{figure}

Figure \ref{fig:efs-example} illustrates how the proposed method performs for recovering eigenfunctions. Here we used binary data, Case 1 eigenfunctions, $I = 1000$, $J = 10$, $K = 100$, $w = 5\%$, and $b_0 = 0$. Results indicate that the proposed method accurately captures the true eigenfunctions at both subject (level~1) and subject-visit (level~2) levels.

\begin{figure}[!tbh]
\centering
\includegraphics[width=0.7\textwidth]{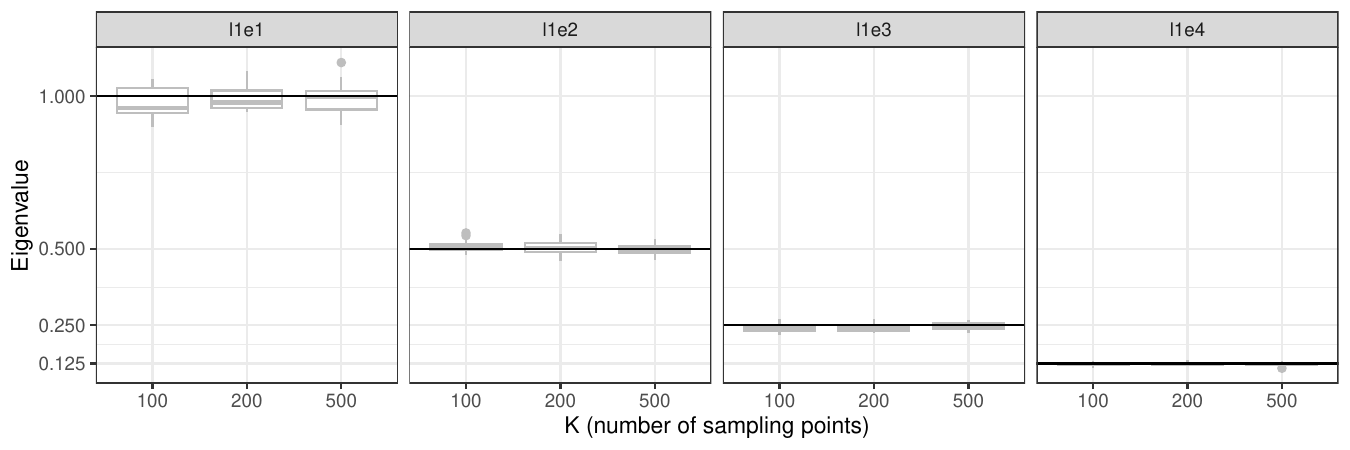}
\includegraphics[width=0.7\textwidth]{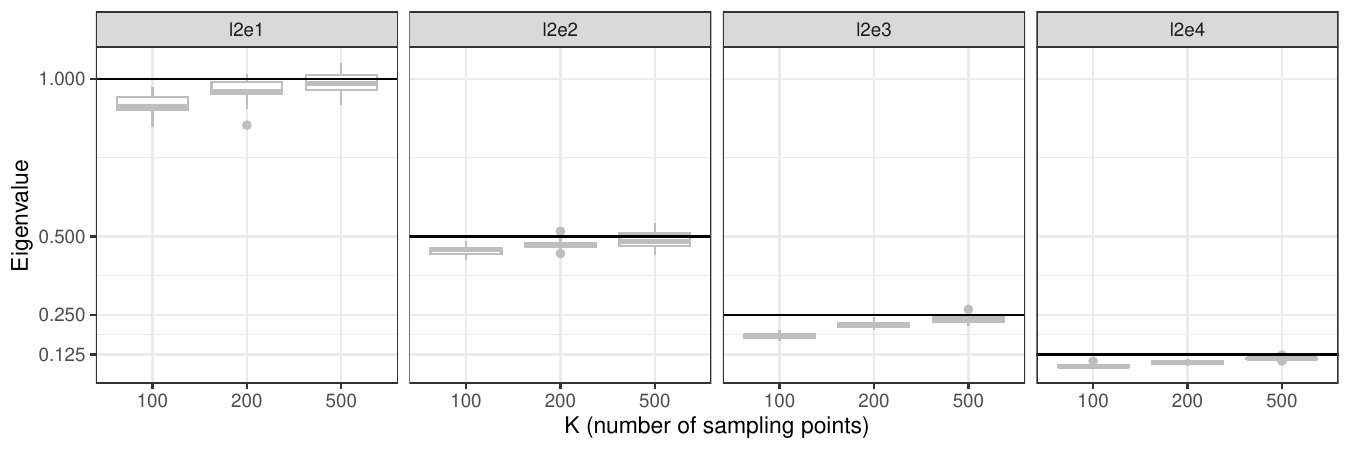}
  \caption{Boxplots of estimated eigenvalues for different numbers of sampling points $K$. Solid black lines are the true eigenvalues. We simulated binary data using Case 1 eigenfunctions, sample size $I = 1000$, number of visits $J = 10$, percentage of data used for constructing local bins $w = 5\%$, and fixed effect intercept $b_0 = 0$.}
  \label{fig:eigenvals-by-S}
\end{figure}

Figure~\ref{fig:eigenvals-by-S} displays the boxplots of the estimated eigenvalues at levels~1 and~2 for different numbers of sampling points $K$, holding other parameters fixed. The solid black lines represent the true eigenvalues. Eigenvalues at level~1 exhibit negligible bias, whereas level~2 eigenvalues show a modest bias that decreases with increasing $K$.

Table~\ref{table:mise-lpscale-by-sample-size} reports the mean squared error (MSE) of the linear predictors for binary and Poisson data under Case~1 and Case~2 basis functions at various sample sizes $I$. As $I$ increases, our method more accurately recovers the latent linear predictors.

\begin{table}[ht]
\begin{center}
\begin{adjustbox}{width=\textwidth}
\begin{tabular}{lcccccccccc}
\toprule
\multicolumn{1}{c}{\,} & \multicolumn{5}{c}{Binary} & \multicolumn{5}{c}{Poisson}\\ \cmidrule(lr){2-6}\cmidrule(lr){7-11}
Eigenfunction & I = 50 & I = 100 & I = 200 & I = 500 & I = 1000 & I = 50 & I = 100 & I = 200 & I = 500 & I = 1000\\ \midrule
Case 1 & 0.292 & 0.281 & 0.271 & 0.268 & 0.267 & 0.042 & 0.034 & 0.031 & 0.029 & 0.028\\
Case 2 & 0.313 & 0.297 & 0.294 & 0.291 & 0.291 & 0.102 & 0.103 & 0.095 & 0.096 & 0.098\\
\bottomrule
\end{tabular}
\end{adjustbox}
\end{center}
\caption{Mean squared error (MSE) of the linear predictors, averaged over 20 runs. We fixed the number of visits $J = 10$, percentage of data used for constructing local bins $w = 5\%$, and fixed effect intercept $b_0 = 0$.}
\label{table:mise-lpscale-by-sample-size}
\end{table}

More simulation results for varying the number of subjects, number of sampling points, number of visits, bin width, and the fixed effect intercepts are provided in Table \ref{supp-table:mise-efs-by-sample-size} to \ref{supp-table:mise-lpscale-by-intercept} in the supplementary materials. 

\subsection{Computation Time} \label{sec:computation-time}
Table~\ref{table:computation-time-sample-size} shows the median computation time (in minutes) over 20 repeated simulations for different sample sizes. Steps~1--3 generally take only a few minutes or less. The primary bottleneck is Step~4, where MCMC is employed to estimate scores using the entire dataset. With $1{,}000$ post warm-up MCMC iterations, the median runtime ranged from 10~minutes to 26~hours for binary data, and from 47~minutes to 6~days for Poisson data, depending on the sample size and eigenfunctions. This also illustrates the scale of our simulations, providing the most extensive such study to date. 

\begin{table}[ht]
\begin{center}
\begin{adjustbox}{width=\textwidth}
\begin{tabular}{llrrrrrr}
\toprule
\multicolumn{1}{c}{\,} & \multicolumn{1}{c}{\,} & \multicolumn{3}{c}{Binary} & \multicolumn{3}{c}{Poisson}\\ \cmidrule(lr){3-5}\cmidrule(lr){6-8}
Eigenfunction & Sample Size & Step 1 and 2 & Step 3 & Step 4 & Step 1 and 2 & Step 3 & Step 4 \\ \midrule
Case 1 & 50 & 0 & 0.002 & 22 & 1 & 0.002 & 47\\
 & 100 & 1 & 0.003 & 60 & 1 & 0.002 & 265\\
& 200 & 2 & 0.003 & 95 & 2 & 0.003 & 544\\
 & 500 & 4 & 0.005 & 543 & 5 & 0.005 & 1424\\
& 1000 & 12 & 0.008 & 1572 & 10 & 0.008 & 3231\\ \midrule 
Case 2 & 50 & 0 & 0.002 & 11 & 1 & 0.002 & 296\\
& 100 & 1 & 0.003 & 78 & 1 & 0.003 & 763\\
 & 200 & 2 & 0.003 & 97 & 2 & 0.003 & 1377\\
& 500 & 4 & 0.005 & 551 & 5 & 0.004 & 3240\\
 & 1000 & 9 & 0.007 & 678 & 11 & 0.007 & 7850\\
\bottomrule
\end{tabular}
\end{adjustbox}
\end{center}
\caption{Median computation time (in minutes) over 20 runs. We fixed the number of visits  $J = 10$, number of sampling points $K = 100$, and percentage of data used for constructing local bins $w = 5\%$. For binary data, we used fixed effect intercept $b_0 = 0$.}
\label{table:computation-time-sample-size}
\end{table}

\section{Application to NHANES Accelerometry Data}\label{sec:real-data}

\subsection{Data Description}
The National Health and Nutrition Examination Survey (NHANES) is a large, ongoing study conducted by the United States Centers for Disease Control and Prevention (CDC) that provides a nationally representative sample of the non-institutionalized US population. NHANES collects data in two-year waves, and wearable accelerometers were deployed in the 2003--2004, 2005--2006, 2011--2012, and 2013--2014 waves. Participants in the 2011--2014 waves wore a wrist-worn ActiGraph GT3X+ accelerometer (ActiGraph, Pensacola, FL) consecutively for nine calendar days, yielding up to seven full days of data (the first and last days are partially observed per study protocol).

Physical activity data provided by NHANES are available in day, hour, and minute resolutions in Monitor Independent Movement Summary (MIMS) units \citep{john2019mims} -- a summary statistic computed from the tri-axial acceleration signal. Specifically, the raw signal is interpolated to 100Hz, extrapolated if the device's dynamic range is exceeded, bandpass-filtered (0.2--5.0Hz), integrated over time to compute an area-under-curve measure for each axis, and then summed across axes to yield a single MIMS value per epoch.

In this paper, we focus on minute-level MIMS data. For data quality control, we define a ``good day of data" as one in which the accelerometer was worn (the wear" flag \verb|PAXPREDM| $\in \{1,2,4\}$) at least $95$\% of the time and contained no data-quality flags (\verb|PAXFLGSM|,=,""). We exclude participants with no good day of data. Out of $14{,}693$ participants who wore an accelerometer, $1{,}090$ were excluded for poor data quality, leaving $13{,}603$ participants eligible.

We link these data to mortality outcomes using death certificate records from the National Center for Health Statistics (NCHS) \citep{mortality2019}, with administrative censoring on December31, 2019. Because the primary focus is on mortality risk in older adults, we further excluded $9{,}149$ participants under age $50$ (age measured at the NHANES interview). We also exclude $9$ individuals whose mortality status is missing. The resulting analytic sample for GM-FPCA consists of $4{,}445$ participants, contributing $28{,}023$ valid days of wear at $1{,}440$ minute-level observations per day. Table,\ref{supp-table:table1} in the supplementary material summarizes demographic characteristics and traditional mortality risk factors. Following \citet{karas2022comparison}, we threshold MIMS at $10.558$ to create a binary active/inactive indicator (active if MIMS$>~10.558$, inactive otherwise).

\subsection{Analysis Results}\label{subsec:results}
We applied GM-FPCA to these minute-level binary activity data using a bin width of $30$ minutes  (roughly $2$\% of the $24$ hour domain).  The local GLMM fitting in Step2 was parallelized because each bin can be processed independently.

\begin{figure}[!tbh]
\centering
\includegraphics[width=0.85\textwidth]{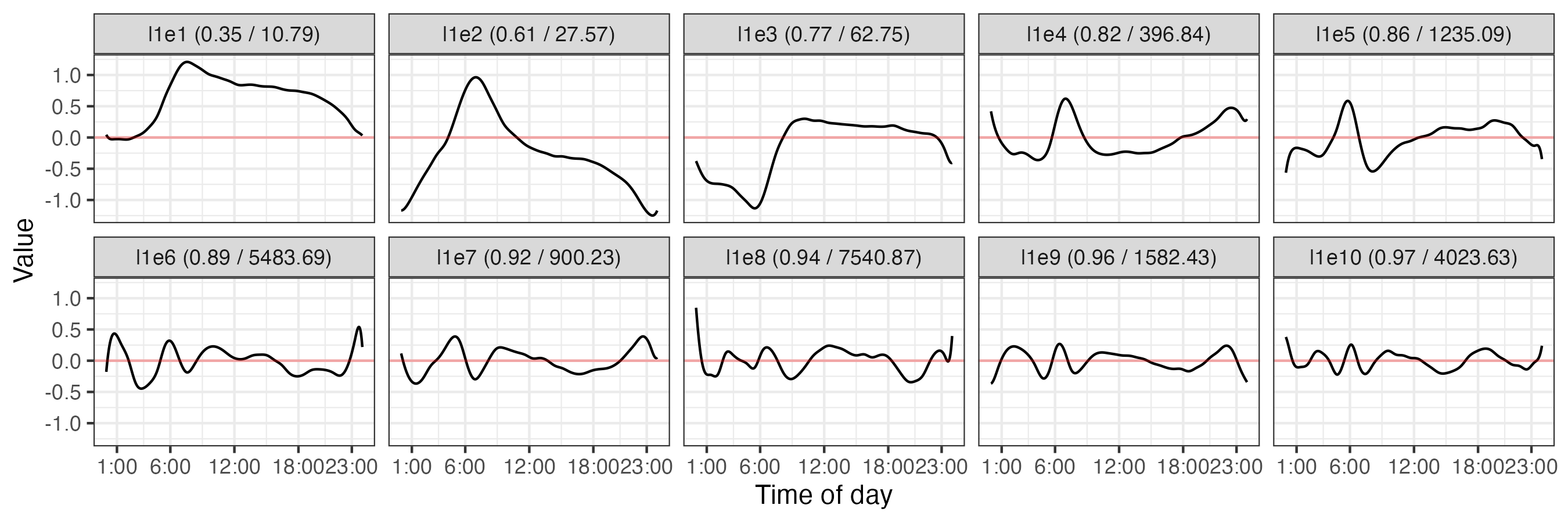}
  \caption{Top 10 level-one eigenfunctions for the binary NHANES activity data. Subplot titles indicate the eigenfunction, cumulative fraction of variance explained within level-one, and the sum of squared second-order differences (SSSOD, see supplementary materials Section \ref{supp-sec:practical-considerations} for the definition). For instance, ``l1e2 (0.61 / 27.57)" denotes the second eigenfunction (l1e2) at level-one, that the first two level-one eigenfunctions together explain $61$\% of the variance at that level, and that the SSSOD is $27.57$.} 
  \label{fig:nhanes-2011-l1ef}
\end{figure}

\begin{figure}[!tbh]
\centering
\includegraphics[width=0.85\textwidth]{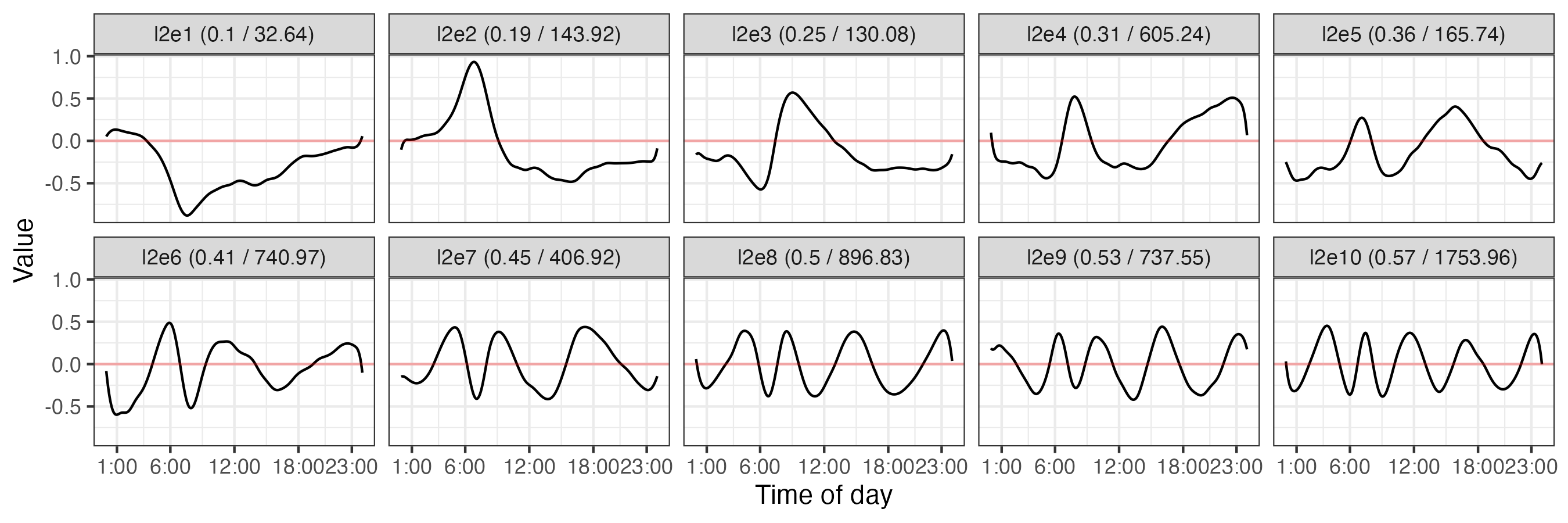}
  \caption{Top ten level-two eigenfunctions from the binary NHANES activity data. Each subplot title gives the eigenfunction identifier, cumulative fraction of variance explained within level-two, and the SSSOD value.}
  \label{fig:nhanes-2011-l2ef}
\end{figure}

Figure~\ref{fig:nhanes-2011-l1ef} plots the top $10$ subject-level (level-one) eigenfunctions, which together explain $47$\% of the total variability in the latent space. Figure~\ref{fig:nhanes-2011-l2ef} shows the top $10$ subject-visit-level (level-two) eigenfunctions, explaining the remaining $53$\%. 

At level-one (Figure~\ref{fig:nhanes-2011-l1ef}), the first three eigenfunctions are relatively easy to interpret. Recall that subject-level (level-one) eigenfunctions capture the main directions of variation of the average within-subject physical activity trajectory. Figure \ref{fig:nhanes-2011-l1ef} indicates that the first eigenfunction is near zero between midnight and $4$AM, then again from $11$PM to midnight, and is positive from $4$AM to $11$PM, peaking around $8$AM. Participants scoring high on this component typically exhibit more daytime activity and less nighttime activity. 

\begin{figure}[!tbh]
\centering
\includegraphics[width=0.3\textwidth]{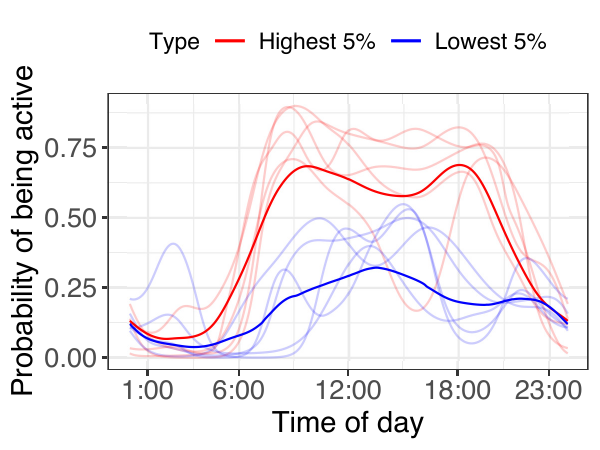}
\includegraphics[width=0.3\textwidth]{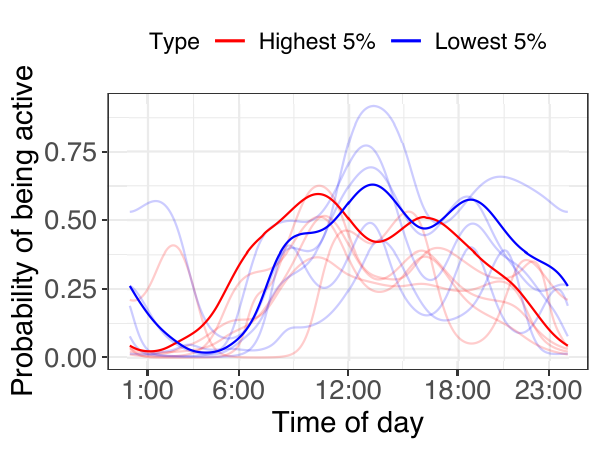}
\includegraphics[width=0.3\textwidth]{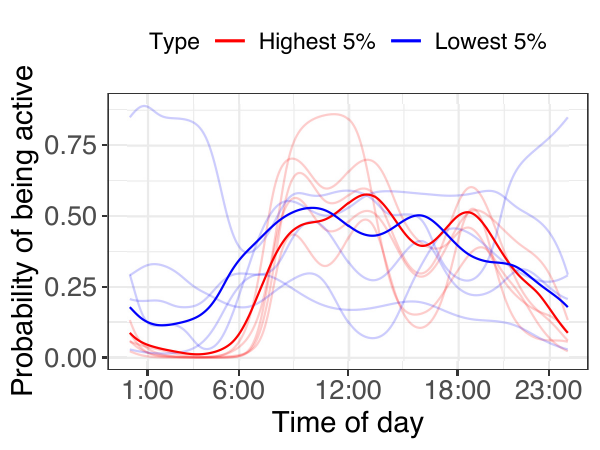}
  \caption{Smoothed daily activity profiles for subjects in the top (red) and bottom (blue) 5\% of scores on level-one eigenfunctions 1--3 (left to right). Thin lines show subject-specific smooths averaged over days from 5 randomly selected subjects; thick lines are group means.}
  \label{fig:ef-high-low}
\end{figure}

To illustrate, Figure~\ref{fig:ef-high-low} shows smoothed profiles for five randomly selected participants in the top and bottom $5$\% of scores for the corresponding eigenfunctions. Smoothing was applied directly to binary activity data for each day using generalized additive models \citep{wood2017gam} and averaged across days. These average activity profiles are shown as thin lines, while the average of all study participants within the $5$\% lowest and highest scores are shown as a thick line of the same color (red for the highest $5$\% and blue for the lowest $5$\%). Figure~\ref{fig:ef-high-low} (left) indicates that participants with high ``l1e1" scores (red) tend to be consistently active during daytime hours, whereas those with low scores (blue) are far more sedentary during the day. 
Similar plots for ``l1e2" and ``l1e3" are displayed in the middle and right panels of Figure \ref{fig:ef-high-low}. Results indicate that participants with higher scores for ``l1e2" tend to be more active in early morning and less active throughout the rest of the day, which is consistent with the shape of ``l1e2'' in Figure \ref{fig:nhanes-2011-l1ef}. We leave the interpretation of the third level-one eigenfunction (``l1e3'') to the reader.

The first three eigenfunctions in level-one capture $77$\% of variability at level one. The remaining level-one eigenfunctions contribute less total variability, but exhibit increasingly complex shapes, measured by the SSSOD statistics. For example, the SSSOD of eigenfunction one is orders of magnitude smaller than those of eigenfunctions $4$-$10$. We do not propose a rigorous analysis of the SSSOD measure at this time, though we found it to be useful. 

Figure~\ref{fig:nhanes-2011-l2ef} displays level-two eigenfunctions. These capture day-to-day deviations from each subject’s average activity pattern. While level-two eigenfunctions explain a similar proportion of total variability as level-one eigenfunctions, the variability within level-two is more spread out across eigenfunctions. For example, the first six level-two eigenfunctions account for only 41\% of the variability within level-two, whereas the first six level-one eigenfunctions explain $89$\% of the variability at that level. The results are consistent with our intuition about physical activity patterns: that the patterns of day-to-day variability are more heterogeneous than those of subject-to-subject variability.

A Bayesian multilevel model~\eqref{eqn:global-glmm} conditioning on the top ten eigenfunctions at each level is used to estimate scores. In spite of dimension reduction efforts, fitting the Bayesian model still raises extraordinary computational challenges. Indeed, we have $4{,}445$ subjects and $28{,}023$ days, thus $4445\times10 + 28023\times10 = 324{,}680$ random-effect parameters. We follow \citet{leroux2023fast} by partitioning the data into $7$ subsets of about $635$ subjects each, greatly reducing the parameter space for each model to roughly $46{,}382$. They showed that this strategy yields results nearly indistinguishable from a single joint fit while remaining computationally tractable. Of course, this could be done jointly, but we suggest balancing what is reasonable with what is truly necessary. 

Finally, we investigate whether the subject-specific scores (level-one) and the standard deviation of subject-visit-specific scores (level-two) are associated with time to death after accounting for traditional mortality risk factors. Of the $4{,}445$ subjects, 127 have missing covariates, leaving $4{,}318$ in the final Cox regression. Table~\ref{table:cox-models} displays results from four Cox models: Model 1 includes only conventional risk factors; Model 2 adds level-one scores; Model 3 additionally includes the day-to-day (level-two) score standard deviations; and Model 4 performs stepwise selection on Model 3 based on the Akaike information criterion (AIC). 

From Model 2, we observe that adding level-one scores substantially improves fit and reveals that more daytime activity (higher scores on eigenfunction 1, hazard ratio $0.66$, $95$\% CI $[0.60, 0.72]$) and less nighttime activity (higher scores on eigenfunction 3, hazard ratio $0.67$, $95$\% CI $[0.60, 0.75]$) reduce mortality risk, whereas being active only in  early morning (eigenfunction 2, hazard ratio $1.18$, $95$\% CI $[1.06, 1.32]$) is associated with increased risk. 

In Model3, certain level-two variability measures also emerge as significant predictors of mortality. For example, the standard deviation of scores for level-two eigenfunction 3 (hazard ratio $0.49$, 95\% CI $[0.27, 0.90]$) and eigenfunction 4 (hazard ratio $2.01$, 95\% CI $[1.08, 3.75]$) are both associated with mortality risk, indicating that day-to-day activity fluctuations at specific times of day may be important.

\begin{figure}[!tbh]
\centering
\includegraphics[width=0.49\textwidth]{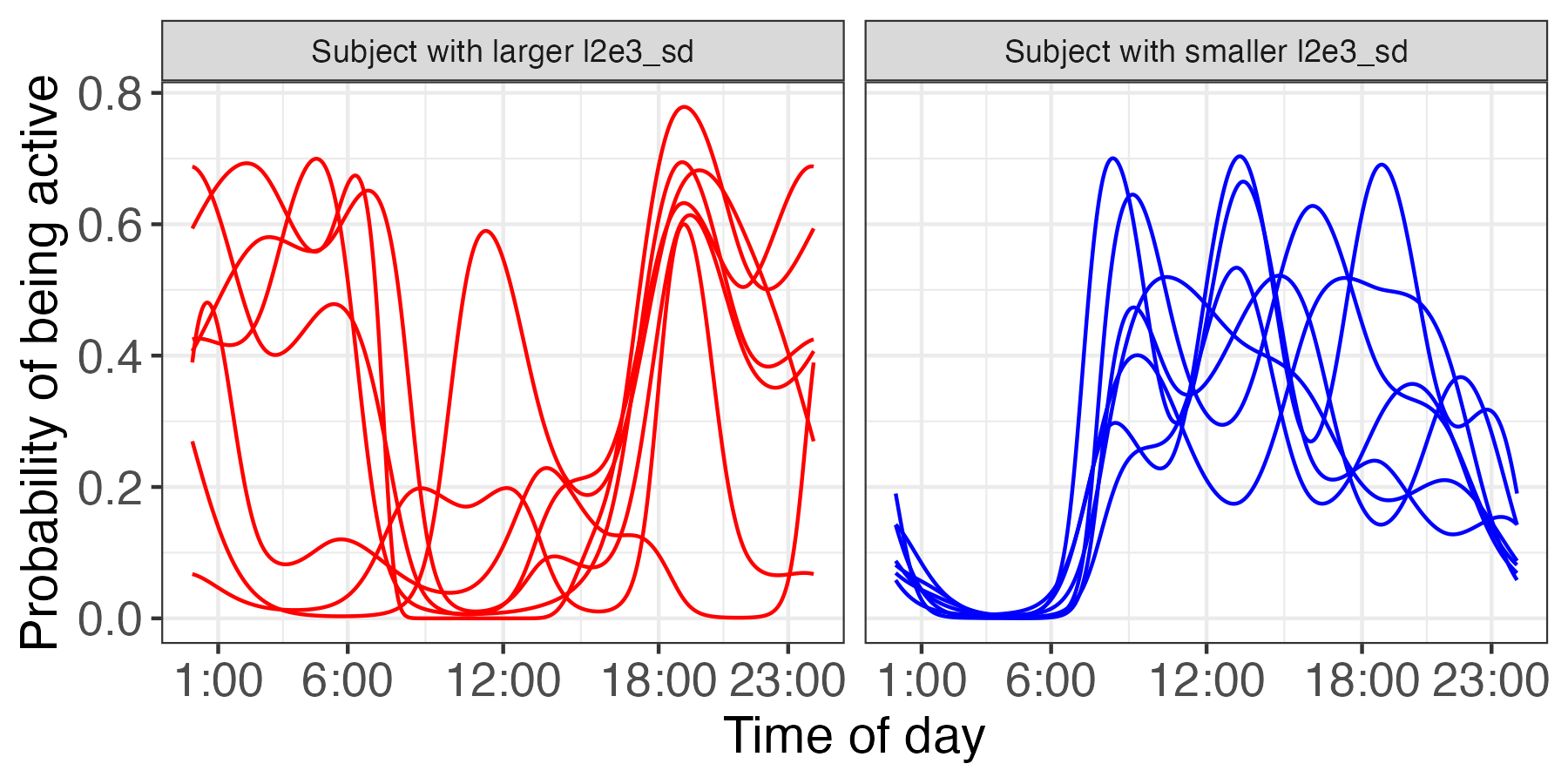}
\includegraphics[width=0.49\textwidth]{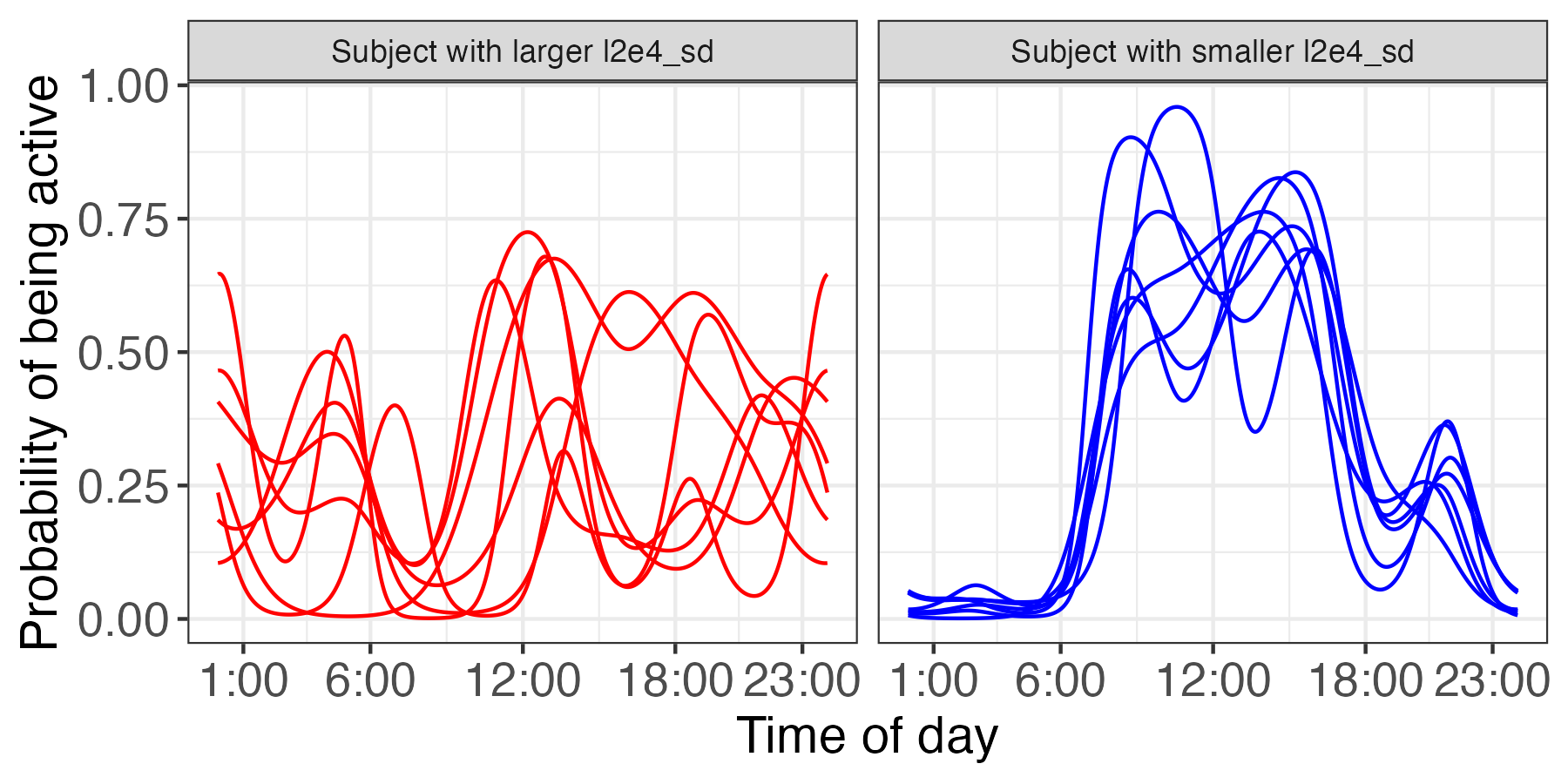}
  \caption{Smoothed daily activity profiles for subjects with larger (red) versus smaller (blue) values of standard deviation of scores of level-two eigenfunction 3 (two plots on the left) and eigenfunction 4 (two plots on the right). Each subplot shows one subject; each line represents a single day.}
  \label{fig:l2esd-high-low}
\end{figure}

To further interpret these findings, we refer to Figure~\ref{fig:nhanes-2011-l2ef}. From the shape of ``l2e3", days with higher scores on this eigenfunction have lower activity at 6 AM and higher activity at 9 AM. Consequently, participants whose ``l2e3" score varies extensively across days (large ``l2e3\_sd") have larger fluctuations between early- and mid-morning. To further illustrate this, Figure~\ref{fig:l2esd-high-low} (left panel) shows smoothed binary activity profiles for two subjects: the one shown in red exhibits larger ``l2e3\_sd'', while the one shown in blue has a smaller value. Each line represents a single day. The fluctuation in day-to-day early morning activity is evident for  the subject with larger values of ``l2e3\_sd'' (red). 

Having a higher score on ``l2e4" is consistent with being more active at 7 AM and 11 PM. The right panel of Figure~\ref{fig:l2esd-high-low} displays a subject with higher ``l2e4\_sd" (red) and another with a lower value (blue). The subject with larger score (red) exhibits larger day-to-day variability at different times of the day, and, especially, at 7 AM and 11 PM. In contrast,  the subject with the lower score (blue) has a more consistent day-to-day schedule. The hazard ratio of $2.01$ for ``l2e4\_sd" suggests that greater variability in early-morning or late-evening activity is associated with elevated mortality risk.

We do not yet have a biological or behavioral explanation for why variability in daily physical activity patterns at particular times is linked to mortality. Further studies in other large accelerometry cohorts (e.g., the UK Biobank) can help validate these findings, address multiple-testing concerns, and explore underlying mechanisms. Although \citet{melin2016variability} reported an association between physical activity variability and mortality, they used scalar summaries. We believe GM-FPCA is the first method to capture time-specific variability throughout the day, providing a more nuanced understanding of how temporal fluctuations in physical activity might influence health outcomes.

\begin{table}[!htbp] \centering 
\footnotesize
\caption{Cox proportional hazards models for time-to-death in NHANES with different sets of covariates.} 
\label{table:cox-models} 
\begin{tabular}{@{\extracolsep{5pt}}lcccc} 
\\[-1.8ex]\hline 
\hline \\[-1.8ex] 
& \multicolumn{4}{c}{Hazard Ratio (95\% CI)}\\ 
\\[-1.8ex] & Model 1 & Model 2 & Model 3 & Model 4\\ 
\hline \\[-1.8ex] 
Age & 1.10$^{***}$ (1.09, 1.11) & 1.09$^{***}$ (1.08, 1.10) & 1.09$^{***}$ (1.08, 1.10) & 1.09$^{***}$ (1.08, 1.10) \\ 
Gender: Female & 0.74$^{***}$ (0.63, 0.86) & 0.83$^{*}$ (0.71, 0.97) & 0.88 (0.75, 1.03) & 0.88 (0.75, 1.03) \\ 
Race & & & & \\
(Ref: Non-Hispanic White) & & & & \\
\;\; Non-Hispanic Black & 0.84 (0.70, 1.02) & 0.81$^{*}$ (0.67, 0.99) & 0.84 (0.69, 1.02) & 0.85 (0.70, 1.03) \\ 
\;\; Mexican American & 0.59$^{**}$ (0.43, 0.82) & 0.65$^{**}$ (0.47, 0.90) & 0.67$^{*}$ (0.48, 0.94) & 0.69$^{*}$ (0.50, 0.96) \\ 
\;\; Non-Hispanic Asian & 0.44$^{***}$ (0.31, 0.63) & 0.44$^{***}$ (0.31, 0.64) & 0.45$^{***}$ (0.31, 0.65) & 0.46$^{***}$ (0.32, 0.66) \\ 
\;\; Other Hispanic & 0.57$^{***}$ (0.41, 0.79) & 0.66$^{*}$ (0.48, 0.92) & 0.67$^{*}$ (0.48, 0.93) & 0.68$^{*}$ (0.49, 0.94) \\ 
\;\; Other Race & 1.22 (0.73, 2.06) & 1.09 (0.65, 1.85) & 1.31 (0.78, 2.21) & 1.34 (0.80, 2.26) \\ 
BMI (Ref: Normal) & & & & \\
\;\; Underweight & 1.69$^{*}$ (1.12, 2.55) & 1.61$^{*}$ (1.06, 2.43) & 1.65$^{*}$ (1.09, 2.52) & 1.58$^{*}$ (1.04, 2.39) \\ 
\;\; Overweight & 0.66$^{***}$ (0.54, 0.80) & 0.63$^{***}$ (0.52, 0.76) & 0.63$^{***}$ (0.52, 0.77) & 0.64$^{***}$ (0.52, 0.77) \\ 
\;\; Obese & 0.70$^{***}$ (0.58, 0.85) & 0.59$^{***}$ (0.49, 0.72) & 0.59$^{***}$ (0.48, 0.72) & 0.60$^{***}$ (0.49, 0.73) \\ 
Overall Health & & & & \\
(Ref: Excellent) & & & & \\
\;\; Very good & 0.76 (0.55, 1.06) & 0.82 (0.59, 1.14) & 0.85 (0.61, 1.18) & 0.82 (0.59, 1.14) \\ 
\;\; Good & 0.66$^{***}$ (0.53, 0.82) & 0.69$^{**}$ (0.55, 0.86) & 0.71$^{**}$ (0.57, 0.90) & 0.70$^{**}$ (0.56, 0.88) \\ 
\;\; Fair & 1.53$^{***}$ (1.27, 1.84) & 1.47$^{***}$ (1.22, 1.77) & 1.46$^{***}$ (1.20, 1.76) & 1.48$^{***}$ (1.23, 1.79) \\ 
\;\; Poor & 2.68$^{***}$ (2.05, 3.50) & 2.45$^{***}$ (1.87, 3.20) & 2.33$^{***}$ (1.78, 3.06) & 2.43$^{***}$ (1.86, 3.17) \\ 
Diabetes (Ref: No) & & & & \\
\;\; Borderline & 0.80 (0.52, 1.23) & 0.84 (0.54, 1.29) & 0.85 (0.55, 1.30) &  \\ 
\;\; Yes & 1.25$^{*}$ (1.05, 1.49) & 1.17 (0.98, 1.39) & 1.16 (0.97, 1.39) &  \\ 
Arthritis: Yes  & 1.04 (0.89, 1.21) & 1.04 (0.89, 1.22) & 1.03 (0.88, 1.20) &  \\ 
CHD: Yes  & 1.50$^{***}$ (1.22, 1.84) & 1.45$^{***}$ (1.18, 1.78) & 1.43$^{***}$ (1.16, 1.75) & 1.44$^{***}$ (1.18, 1.77) \\ 
Stroke: Yes & 1.42$^{**}$ (1.15, 1.77) & 1.32$^{*}$ (1.06, 1.64) & 1.31$^{*}$ (1.05, 1.64) & 1.32$^{*}$ (1.06, 1.64) \\ 
Cancer: Yes & 1.21$^{*}$ (1.02, 1.44) & 1.24$^{*}$ (1.04, 1.47) & 1.22$^{*}$ (1.03, 1.46) & 1.23$^{*}$ (1.04, 1.47) \\ 
l1e1 &  & 0.66$^{***}$ (0.60, 0.72) & 0.60$^{***}$ (0.54, 0.67) & 0.60$^{***}$ (0.54, 0.66) \\ 
l1e2 &  & 1.18$^{**}$ (1.06, 1.32) & 1.20$^{**}$ (1.07, 1.35) & 1.22$^{***}$ (1.09, 1.36) \\ 
l1e3 &  & 0.67$^{***}$ (0.60, 0.75) & 0.66$^{***}$ (0.59, 0.74) & 0.67$^{***}$ (0.60, 0.75) \\ 
l1e4 &  & 1.00 (0.79, 1.26) & 1.01 (0.80, 1.28) &  \\ 
l1e5 &  & 1.00 (0.82, 1.20) & 0.99 (0.81, 1.21) &  \\ 
l1e6 &  & 0.81 (0.63, 1.05) & 0.80 (0.62, 1.04) & 0.81 (0.64, 1.02) \\ 
l1e7 &  & 0.98 (0.77, 1.24) & 0.97 (0.76, 1.23) &  \\ 
l1e8 &  & 1.07 (0.78, 1.48) & 1.14 (0.83, 1.59) &  \\ 
l1e9 &  & 1.44$^{*}$ (1.09, 1.90) & 1.39$^{*}$ (1.04, 1.84) & 1.38$^{*}$ (1.04, 1.82) \\ 
l1e10 &  & 0.87 (0.62, 1.21) & 0.87 (0.62, 1.23) &  \\ 
l2e1\_sd &  &  & 1.11 (0.75, 1.64) &  \\ 
l2e2\_sd &  &  & 0.82 (0.51, 1.33) &  \\ 
l2e3\_sd &  &  & 0.49$^{*}$ (0.27, 0.90) & 0.50$^{*}$ (0.29, 0.86) \\ 
l2e4\_sd &  &  & 2.01$^{*}$ (1.08, 3.75) & 1.99$^{*}$ (1.11, 3.56) \\ 
l2e5\_sd &  &  & 0.98 (0.53, 1.84) &  \\ 
l2e6\_sd &  &  & 1.24 (0.72, 2.14) &  \\ 
l2e7\_sd &  &  & 0.62 (0.31, 1.22) & 0.61 (0.32, 1.17) \\ 
l2e8\_sd &  &  & 0.75 (0.36, 1.56) &  \\ 
l2e9\_sd &  &  & 1.37 (0.66, 2.85) &  \\ 
l2e10\_sd &  &  & 0.95 (0.44, 2.03) &  \\ 
\hline 
Concordance & 0.788 & 0.805 & 0.807 & 0.805\\
\hline 
Integrated Brier score & 0.071 & 0.067 & 0.067 & 0.067 \\
\hline 
\hline \\[-1.8ex] 
\textit{Note:}  & \multicolumn{2}{r}{$^{*}$p$<$0.05; $^{**}$p$<$0.01; $^{***}$p$<$0.001} \\ 
\end{tabular} 
\end{table}

\section{Discussion}\label{sec:discussion}
Motivated by the need to analyze large binary multilevel physical activity data, we proposed GM-FPCA, a novel method for generalized (non-Gaussian) multilevel functional outcomes. Through binning and fitting local GLMMs, GM-FPCA efficiently extracts local random-effect estimates for subsequent decomposition. While our study was motivated by binary data, the framework can accommodate other link functions and non-Gaussian functional outcomes such as categorical or count. 

Extensive simulation experiments confirm GM-FPCA’s ability to accurately recover eigenfunctions and scores across diverse sample sizes, numbers of visits, and numbers of sampling points. 
Its scalability is demonstrated by both the simulation studies and an application to the 2011--2014 NHANES accelerometry data, which include $4{,}445$ subjects, $28{,}023$ days of wear, and $1{,}440$ minute-level observations per day.

We illustrated how GM-FPCA identified interpretable activity patterns associated with mortality in binary NHANES data and, for the first time, uncovered a potential link between mortality and variability in daily activity at particular times of day. These insights carry public-health relevance and may inform individualized lifestyle interventions. 

Despite GM-FPCA’s flexibility, we acknowledge that Step4 can still be computationally intensive for very large datasets. Downsampling the domain in Step 4 (see supplementary materials for preliminary findings) offers a potential compromise between accuracy and runtime, and further research is needed to refine this approach.   
Moreover, while we focused on modeling subject- and subject-visit-specific patterns, incorporating covariates directly in Step 2 could help separate known sources of variability.

In conclusion, GM-FPCA provides a flexible, robust, and interpretable method for analyzing non-Gaussian multilevel functional data. Our results underscore how leveraging the inherent structure and rich information in large accelerometry datasets can lead to valuable scientific and clinical insights, guiding future research and interventions. The \texttt{R} code for GM-FPCA is provided in the supplementary materials to facilitate adoption and further methodological extensions.

\section*{Acknowledgement}
The authors gratefully acknowledge use of the facilities at the Joint High Performance Computing Exchange (JHPCE) in the Department of Biostatistics, Johns Hopkins Bloomberg School of Public Health that have contributed to the research results reported within this paper.

%%%%%%%%%%%%%%%%%
%%% Supplemental %%%
%%%%%%%%%%%%%%%%%%
\clearpage

\section*{Supplementary Material}
\setcounter{page}{1}
\setcounter{section}{0}
\setcounter{table}{0}
\setcounter{figure}{0}

\renewcommand{\thesection}{S.\arabic{section}}
\renewcommand{\thetable}{S.\arabic{table}}
\renewcommand{\theequation}{S.\arabic{equation}}
\renewcommand{\thefigure}{S.\arabic{figure}}

\section{Implementation}
The following code (Listing~\ref{code:gmfpca}) demonstrates how to implement Steps~1--3 of the GM-FPCA algorithm for binary outcomes. 
\lstset{language=R,
    basicstyle=\small\ttfamily,
    commentstyle=\color{blue},
    otherkeywords={0,1,2,3,4,5,6,7,8,9},
    morekeywords={TRUE,FALSE},
    deletekeywords={data,frame,length,as,character,_ls},
    keywordstyle=\color{black},
    caption={\texttt{R} code for the GM-FPCA algorithm (Steps~1--3) for binary data.},
    % captionpos=b,     
    frame=TB,
    numbers=left,                    
    numbersep=5pt,                   
    numberstyle=\small\color{gray},
    label={code:gmfpca}
}
\begin{lstlisting}[language=R]
library(tidyverse)
library(lme4)
library(refund)
dat <- read.csv("...")
fit_ls  <- vector(mode = "list", length = S+1)  
for(j in 0:S){  
  # Step 1: Create data bins.
  sind_j <- (j - ceiling(bin_len/2)):(j + ceiling(bin_len/2)) %% (S+1)
  df_j <- df %>% filter(time %in% sind_j)
  
  # Step 2: Fit local GLMMs; extract linear predictor estimates.
  fit_j <- glmer(
    value ~ 1 + visit + (1|id) + (1|id:visit),
    data = df_j, family = binomial,
  )
  fit_ls[[j+1]] <- list(
    "id_visit" = rownames(coef(fit_j)$"id:visit"),
    "eta_i" = rep(coef(fit_j)$"id"[[1]], each = J) + 
      coef(fit_j)$"id:visit"[[1]],
    "time" = j
  )
}

# Step 3: Run MFPCA
fit_df <- bind_rows(fit_ls)
id_visit <- strsplit(fit_df$id_visit, ":")
fit_df$id <- sapply(id_visit, function(x) as.integer(x[1]))
fit_df$visit <- sapply(id_visit, function(x) as.integer(x[2]))
fit_df <- arrange(fit_df, id, visit, time)
fit_df_wide <- pivot_wider(
    select(fit_df, c(id, visit, time, eta_i)), 
    id_cols = c(id, visit), names_from = time, 
    values_from = eta_i, names_prefix = "T"
)
mfpca.bin <- mfpca.face(
    as.matrix(select(fit_df_wide, -c(id, visit))), 
    id = fit_df_wide$id,
    visit = fit_df_wide$visit,
    argvals = 0:S, 
    pve = 0.95
  )
\end{lstlisting}
\newpage
Listing~\ref{code:step4} provides the \texttt{Stan} implementation for Step~4 of the algorithm, focused on binary data. Minor modifications are needed for Poisson or other exponential-family outcomes.
\lstset{
    basicstyle=\small\ttfamily,
    commentstyle=\color{blue},
    otherkeywords={0,1,2,3,4,5,6,7,8,9},
    morekeywords={TRUE,FALSE},
    deletekeywords={data,frame,length,as,character,_ls},
    keywordstyle=\color{black},
    caption={\texttt{Stan} code for Step~4 of the GM-FPCA algorithm under binary data.}
    % captionpos=b,     
    frame=TB,
    numbers=left,                    
    numbersep=5pt,                   
    numberstyle=\small\color{gray},
    label={code:step4}
}
\begin{lstlisting}
functions {
  real partial_sum_lpmf(array[] int y, int start, int end, vector eta) {
    return bernoulli_logit_lupmf(y | eta[start:end]);
  }
}
data {
  int<lower=0> Nsubs;    // number of individuals 
  int<lower=0> Nfuns; // number of obserrved functions
  int<lower=0> K;    // number of observations per function
  int<lower=0> L;    // number of level 1 eigenfunctions
  int<lower=0> M;    // number of level 2 eigenfunctions
  int<lower=0> ylen;    // Length of the outcome vector
  array[ylen]  int        visit;       // a vector of the visit variable
  array[ylen] int<lower=0,upper=1> y; // outcome (binary functional data)
  array[Nsubs] int nvisits; // number of repeated functions per subject
  row_vector[K] mu_0; // estimated intercept function
  matrix[K, max(nvisits)] mu_j;    // visit specific shift 
  matrix[L,K]   efuncs_l1; // level 1 eigen functions
  matrix[M,K]   efuncs_l2; // level 2 eigen functions
  int<lower=1>  grainsize;
}
parameters {
  matrix[Nsubs, L] xil1;   // level 1 random effects
  matrix[Nfuns, M] xil2;   // level 2 random effects
  vector<lower=0>[L] sigmal1sq; // sd of the random intercept term
  vector<lower=0>[M] sigmal2sq; // sd of the random intercept term
}
transformed parameters {
  matrix[Nsubs, L] xil1_sc;   // level 1 random effects
  matrix[Nfuns, M] xil2_sc;   // level 2 random effects
  for(l in 1:L){
      xil1_sc[,l] = xil1[,l] * sqrt(sigmal1sq[l]);
  }
  for(m in 1:M){
      xil2_sc[,m] = xil2[,m] * sqrt(sigmal2sq[m]);
  }
}
model {
  int inx;
  vector[ylen] eta_vec;
  matrix[Nsubs, K] eta_mat_i;
  matrix[Nfuns, K] eta_mat_ij;
  
  // priors
  to_vector(xil1) ~ normal(0,1);
  to_vector(xil2) ~ normal(0,1);
  sigmal1sq ~ inv_gamma(1, 1);
  sigmal2sq ~ inv_gamma(1, 1);
  
  eta_mat_i = xil1_sc * efuncs_l1;
  for(i in 1:Nsubs){
     eta_mat_i[i, ] = eta_mat_i[i, ] + mu_0;
  }
  eta_mat_ij = xil2_sc * efuncs_l2;
  inx = 1;
  for(i in 1:Nsubs){
          for(j in 1:nvisits[i]){
                eta_mat_ij[inx,] =  
                (mu_j[, visit[inx]])' + eta_mat_ij[inx,] + eta_mat_i[i,];
    inx = inx + 1;
    }
  }
  eta_vec = to_vector((eta_mat_ij)');
  
  // likelihood
  target += reduce_sum(partial_sum_lpmf, y, grainsize, eta_vec);
}
\end{lstlisting}
\newpage

\section{Practical Considerations}\label{supp-sec:practical-considerations}
\subsection{Choosing the Bin Width}
In practice, if the bin width is too small, there may not be sufficient repeated observations per subject to reliably estimate local random effects. Conversely, if bins are too wide, local random effect estimates $\widehat{a}_{ik}^*$ and $\widehat{b}_{ijk}^*$ may estimate $a_{i}(s_k)$ and $b_{ij}(s_k)$ poorly, especially when the functional random effects exhibit substantial oscillation relative to the sampling density. From a practical standpoint, it is rarely necessary to identify a single ``best'' bin width. Indeed, multiple reasonable choices often yield similar results, and we recommend trying a few different bin widths, such as 2\% and 5\% of the domain length, to assess consistency. 

\subsection{Choosing the Number of Eigenfunctions} 
The most common approach for choosing the number of eigenfunctions is based on the cumulative percentage of variance explained. While effective, this strategy can sometimes yield more eigenfunctions than necessary, slowing down Step 4 and leading to eigenfunctions with excessive oscillation. To address this, we recommend also considering the sum of squared second-order differences (SSSOD), a discrete analog of the integrated squared second-order derivative (ISSOD). A higher SSSOD indicates greater oscillation and thus reduced interpretability.
\begin{eqnarray*}
    \text{ISSOD} &=& \int_S [f''(s)]^2 ds \;,\\
    \text{SSSOD} &=& K^2\sum_{k = 1}^{K-2} [f(s_{k+2}) - 2f(s_{k+1}) + f(s_k)]^2\;.
\end{eqnarray*}
Focusing on eigenfunctions that balance high variance-explained with lower SSSOD values can yield smoother, more interpretable components while reducing computational complexity in Step 4.

\section{More on Step 4}
\subsection{Prior Distribution} \label{supp-sec:prior-dist}
\begin{figure}[!tbh]
\centering
  \includegraphics[width=\textwidth]{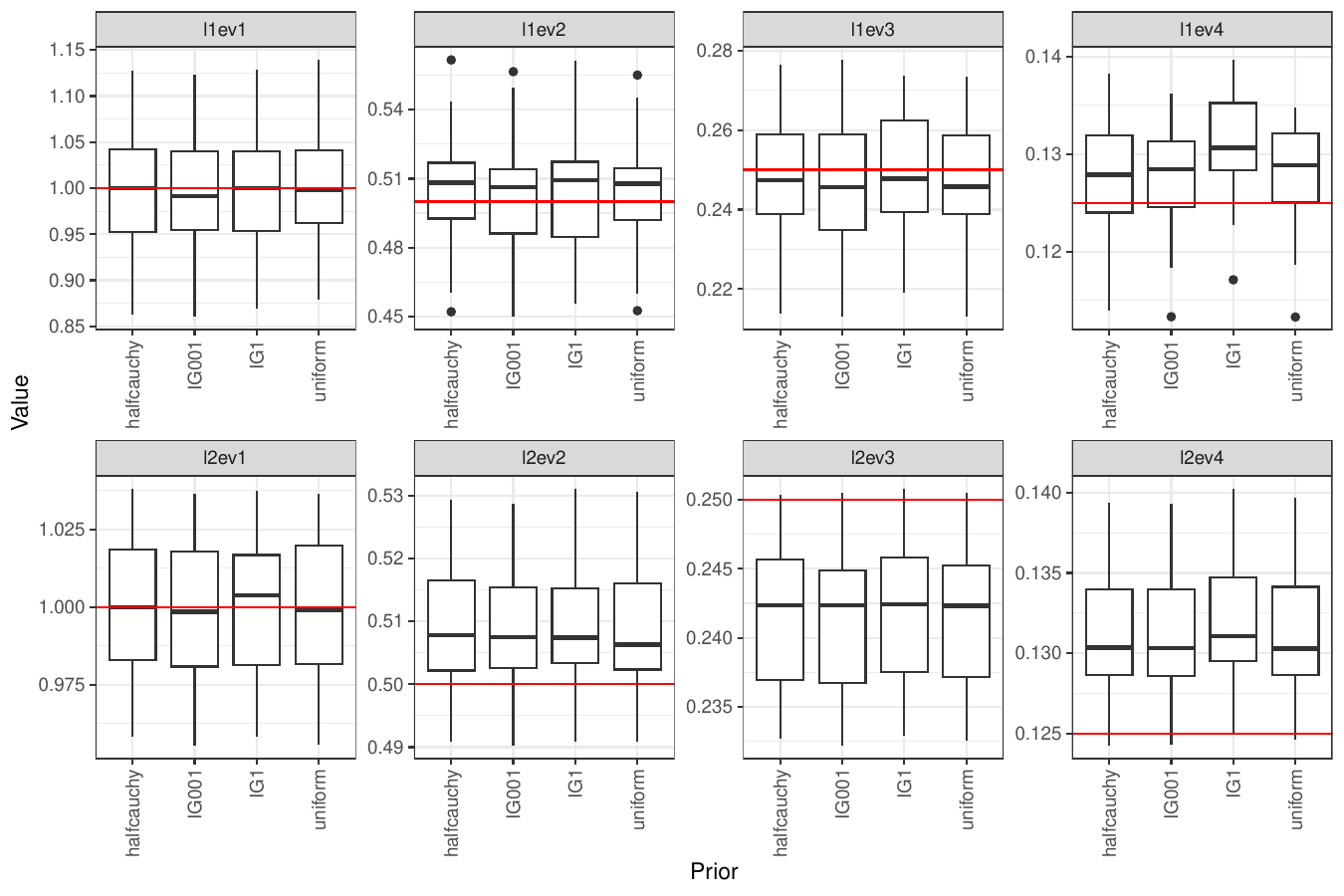}
  \caption{Boxplots of the posterior means of eigenvalues over 20 replicates under different prior distributions. Horizontal red lines indicate the true values.}
  \label{supp-fig:compare-priors}
\end{figure}

We conducted a sensitivity analysis to investigate whether the eigenvalues are sensitive to different prior distributions for the variance parameters, $\sigma_l^2$ and $\sigma_m^2$, in our Bayesian model. Focusing on binary data with $I = 500$ subjects, $J = 10$ visits, $K = 100$ sampling points, 5\% bin width, and intercept $b_0 = 0$, we compared four priors for the random-effect variances. 
Besides $\text{Inv-Gamma}(1,1)$, which was used in simulation and real data analysis, we also considered $\text{Inv-Gamma}(0.001, 0.001)$, $\text{Uniform}(0, 20)$, and $\text{Half-Cauchy}(0, 10)$, 
the last two following \citet{Gelman2006VariancePrior}. Figure~\ref{supp-fig:compare-priors} displays boxplots of the posterior means of the eigenvalues over 20 replicates. The results suggest that the estimated eigenvalues are not sensitive to the choice of prior. We repeated the experiment with $J=2$ visits instead of 10 and observed similar findings. A possible explanation for this ``insensitivity'' lies in the discussion by \citet{Crainiceanu2005WinBUGS}. Let $B_b$ denote the scale parameter for the inverse-gamma prior, and let $\mathbf{b}$ be the corresponding score vector. \citet{Crainiceanu2005WinBUGS} shows that as long as $B_b$ remains small compared to $\|\mathbf{b}\|^2/2$, the posterior distribution is minimally affected by the choice of $B_b$.

\subsection{Downsampling in Step 4}
This section explores downsampling along the functional domain in Step 4 to speed up the estimation of scores. We emphasize that downsampling is not needed in Step 2 because fitting GLMMs in local bins is fully parallelizable; it is also not needed in Step 3 because MFPCA \citep{cui2022mfpca} is highly scalable. 

\begin{figure}[!th]
  \centering
  \includegraphics[width=0.75\textwidth]{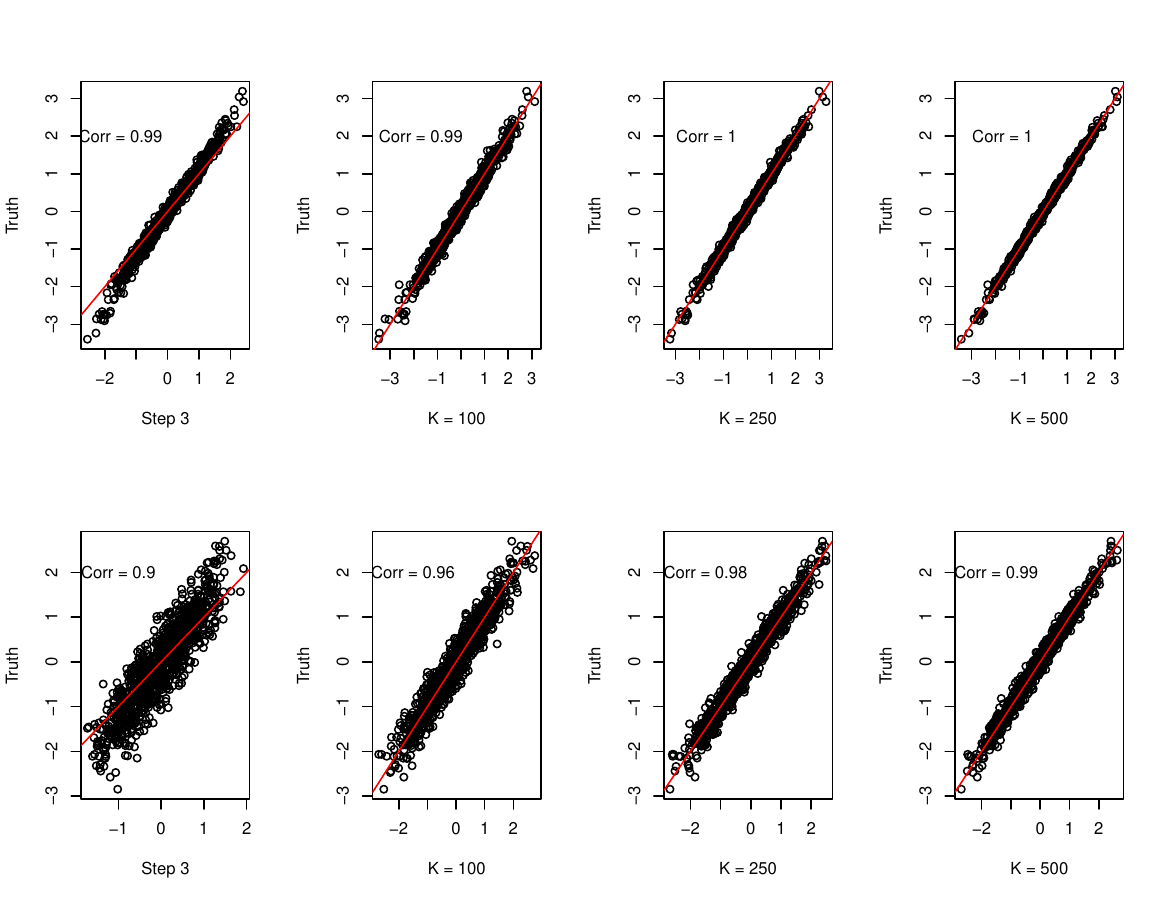}
  \includegraphics[width=0.75\textwidth]{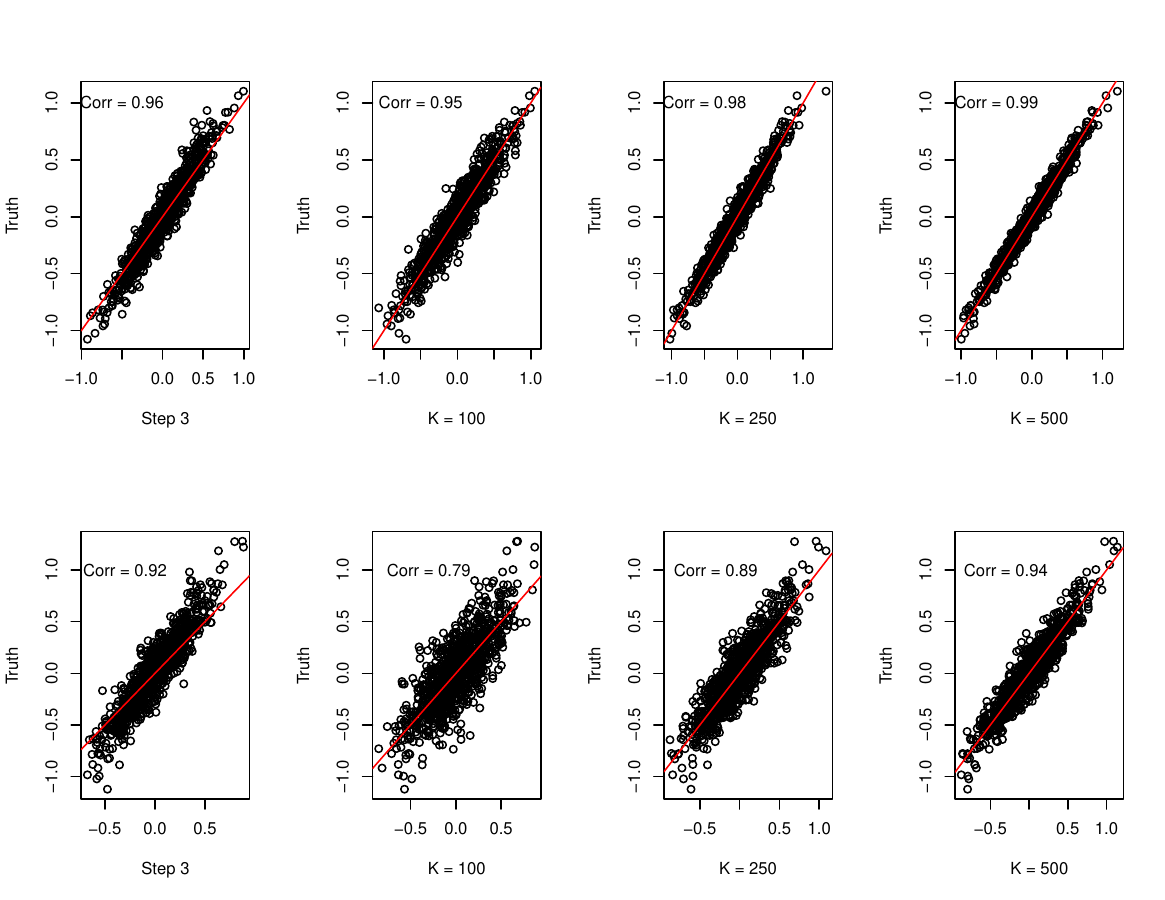}
  \caption{True scores versus those estimated using all sampling points (K=500) and using a subset of points (K=100, 200). Scores obtained from Step 3 are added for comparison. Top to bottom rows: l1e1, l2e1, l1e4, l2e4.}
  \label{supp-fig:downsampling}
\end{figure}

We use a simulated data set so that we know the ground truth. The data set was simulated from $I = 1000$, $J = 5$, $K = 500$, $\mu_0 = 0$, and the response variable is binary. We used $2\%$ of the data to construct local bins. After Step 3, we have the estimated eigenfunctions at all 500 sampling points. To downsample, we take 100 and 250 equally spaced sampling points from the 500 points, and estimate scores using only eigenfunctions and binary responses at the subset of points. Figure \ref{supp-fig:downsampling} shows the estimated scores against the truth for l1e1 (level-one eigenfunction 1), l2e1, l1e4, and l2e4. Scores from Step 3 are also included for comparison. We can see that for top eigenfunctions (l1e1 and l2e1), even downsampling at a rate of 20\% (i.e., taking only 100 sampling points) does not drastically lower the quality of the scores and still provides clear debiasing benefit compared to scores obtained from Step 3. For bottom eigenfunctions (l1e4 and l2e4), however, a higher downsampling rate such as 50\% is needed to secure reasonable score estimates. This is particularly true for level 2 eigenfunctions. In summary, if one is only interested in using the top one or two  eigenfunctions for downstream analysis, we would recommend an aggressive downsampling rate in Step 4 to significantly cut computational cost. 

\newpage
\section{More simulation results}

\subsection{Effect of Sample Size}
We investigate how increasing the number of study participants affects estimation accuracy while keeping other parameters fixed at the following values: number of visits $J = 10$, number of sampling points $K = 100$, and percentage of data used for constructing local bins $w = 5\%$. For binary data, we set the fixed-effect intercept to $b_0 = 0$. Table~\ref{supp-table:mise-efs-by-sample-size} presents the integrated squared error (ISE) of eigenfunctions for different data distributions and basis functions across various sample sizes. The results indicate that ISE decreases rapidly as the number of study participants increases. 
\begin{table}[ht]
\begin{center}
\begin{adjustbox}{width=\textwidth}
\begin{tabular}{llcccccccccc}
\toprule
\multicolumn{1}{c}{\,} & \multicolumn{1}{c}{\,} & \multicolumn{5}{c}{Binary} & \multicolumn{5}{c}{Poisson}\\ \cmidrule(lr){3-7}\cmidrule(lr){8-12}
Basis function & Eigenfunction & I = 50 & I = 100 & I = 200 & I = 500 & I = 1000 & I = 50 & I = 100 & I = 200 & I = 500 & I = 1000\\ \midrule
Case 1 & l1e1 & 0.072 & 0.042 & 0.017 & 0.009 & 0.006 & 0.083 & 0.027 & 0.024 & 0.012 & 0.007\\
 & l1e2 & 0.127 & 0.063 & 0.022 & 0.017 & 0.008 & 0.115 & 0.056 & 0.035 & 0.015 & 0.009\\
 & l1e3 & 0.173 & 0.072 & 0.046 & 0.031 & 0.022 & 0.190 & 0.083 & 0.032 & 0.027 & 0.021\\
 & l1e4 & 0.215 & 0.084 & 0.055 & 0.030 & 0.024 & 0.180 & 0.062 & 0.028 & 0.025 & 0.022\\
 & l2e1 & 0.052 & 0.042 & 0.044 & 0.034 & 0.041 & 0.035 & 0.039 & 0.041 & 0.037 & 0.036\\
 & l2e2 & 0.080 & 0.054 & 0.053 & 0.039 & 0.044 & 0.037 & 0.042 & 0.042 & 0.036 & 0.035\\
 & l2e3 & 0.110 & 0.092 & 0.082 & 0.084 & 0.080 & 0.077 & 0.073 & 0.068 & 0.068 & 0.066\\
 & l2e4 & 0.133 & 0.111 & 0.097 & 0.092 & 0.088 & 0.081 & 0.077 & 0.071 & 0.071 & 0.069\\ 
 \midrule Case 2 & l1e1 & 0.051 & 0.043 & 0.012 & 0.005 & 0.004 & 0.066 & 0.051 & 0.013 & 0.005 & 0.003\\
 & l1e2 & 0.114 & 0.066 & 0.025 & 0.017 & 0.009 & 0.123 & 0.067 & 0.022 & 0.008 & 0.005\\
 & l1e3 & 0.160 & 0.067 & 0.028 & 0.021 & 0.010 & 0.160 & 0.062 & 0.027 & 0.014 & 0.011\\
 & l1e4 & 0.189 & 0.101 & 0.050 & 0.029 & 0.015 & 0.130 & 0.051 & 0.018 & 0.009 & 0.007\\
 & l2e1 & 0.035 & 0.023 & 0.019 & 0.015 & 0.014 & 0.021 & 0.005 & 0.003 & 0.002 & 0.002\\
 & l2e2 & 0.045 & 0.023 & 0.023 & 0.016 & 0.015 & 0.022 & 0.012 & 0.004 & 0.003 & 0.002\\
 & l2e3 & 0.058 & 0.042 & 0.034 & 0.030 & 0.027 & 0.019 & 0.019 & 0.010 & 0.009 & 0.007\\
 & l2e4 & 0.073 & 0.049 & 0.043 & 0.036 & 0.034 & 0.029 & 0.022 & 0.018 & 0.018 & 0.016\\
\bottomrule
\end{tabular}
\end{adjustbox}
\end{center}
\caption{ISE of eigenfunctions averaged over 20 repeats for different sample sizes. ``l1e1" represents the first eigenfunction at level 1, with analogous notation for other eigenfunctions.}
\label{supp-table:mise-efs-by-sample-size}
\end{table}

\subsection{Effect of the Number of Visits}
We now examine the impact of varying the number of visits while keeping other parameters fixed at sample size $I = 1000$, number of sampling points $K = 100$, and percentage of data used for constructing local bins $w = 5\%$. For binary data, we again set the fixed-effect intercept to $b_0 = 0$. Table~\ref{supp-table:mise-efs-by-nvisits} and Table~\ref{supp-table:mise-lpscale-by-nvisits} summarize the ISE of eigenfunctions and the mean squared error (MSE) of linear predictors for different numbers of visits. The results suggest that both ISE and MSE improve as the number of visits increases.
\begin{table}[ht]
\begin{center}
\begin{adjustbox}{width=0.7\textwidth}
\begin{tabular}{llcccccc}
\toprule
\multicolumn{1}{c}{\,} & \multicolumn{1}{c}{\,} & \multicolumn{3}{c}{Binary} & \multicolumn{3}{c}{Poisson}\\ \cmidrule(lr){3-5}\cmidrule(lr){6-8}
Basis function & Eigenfunction & J =  2 & J =  5 & J =  10 & J =  2 & J =  5 & J =  10\\ \midrule
Case 1 & l1e1 & 0.011 & 0.007 & 0.006 & 0.010 & 0.006 & 0.007\\
 & l1e2 & 0.026 & 0.009 & 0.008 & 0.013 & 0.010 & 0.009\\
 & l1e3 & 0.208 & 0.023 & 0.022 & 0.025 & 0.022 & 0.021\\
 & l1e4 & 1.726 & 0.091 & 0.024 & 0.070 & 0.026 & 0.022\\
 & l2e1 & 0.039 & 0.044 & 0.041 & 0.049 & 0.035 & 0.036\\
 & l2e2 & 0.053 & 0.048 & 0.044 & 0.050 & 0.034 & 0.035\\
 & l2e3 & 0.098 & 0.079 & 0.080 & 0.073 & 0.070 & 0.066\\
 & l2e4 & 0.118 & 0.089 & 0.088 & 0.079 & 0.074 & 0.069\\
\midrule Case 2 & l1e1 & 0.009 & 0.004 & 0.004 & 0.007 & 0.005 & 0.003\\
 & l1e2 & 0.040 & 0.012 & 0.009 & 0.008 & 0.006 & 0.005\\
 & l1e3 & 0.398 & 0.014 & 0.010 & 0.019 & 0.010 & 0.011\\
 & l1e4 & 1.563 & 0.401 & 0.015 & 0.281 & 0.011 & 0.007\\
 & l2e1 & 0.024 & 0.017 & 0.014 & 0.004 & 0.001 & 0.002\\
 & l2e2 & 0.028 & 0.018 & 0.015 & 0.006 & 0.002 & 0.002\\
 & l2e3 & 0.041 & 0.029 & 0.027 & 0.010 & 0.007 & 0.007\\
 & l2e4 & 0.047 & 0.036 & 0.034 & 0.019 & 0.017 & 0.016\\
\bottomrule
\end{tabular}
\end{adjustbox}
\end{center}
\caption{ISE of eigenfunctions averaged over 20 repeats for different numbers of visits.}
\label{supp-table:mise-efs-by-nvisits}
\end{table}

\begin{table}[ht]
\begin{center}
\begin{adjustbox}{width=0.7\textwidth}
\begin{tabular}{lcccccc}
\toprule
\multicolumn{1}{c}{\,} & \multicolumn{3}{c}{Binary} & \multicolumn{3}{c}{Poisson}\\ \cmidrule(lr){2-4}\cmidrule(lr){5-7}
Basis function & J =  2 & J =  5 & J =  10 & J =  2 & J =  5 & J =  10\\ \midrule
Case 1 & 0.467 & 0.298 & 0.267 & 0.057 & 0.035 & 0.028\\
Case 2 & 0.398 & 0.312 & 0.291 & 0.120 & 0.097 & 0.098\\
\bottomrule
\end{tabular}
\end{adjustbox}
\end{center}
\caption{MSE of linear predictors averaged over 20 repeats for different numbers of visits.}
\label{supp-table:mise-lpscale-by-nvisits}
\end{table}

\newpage
\subsection{Effect of Bin Width}
In this experiment, we examine the impact of bin width on estimation accuracy while keeping other parameters fixed: sample size $I = 1000$, number of visits $J = 10$, and number of sampling points $K = 100$. For binary data, we set the fixed-effect intercept to $b_0 = 0$. Tables \ref{supp-table:mise-efs-by-binwidth} and \ref{supp-table:mise-lpscale-by-binwidth} report the integrated squared error (ISE) of eigenfunctions and the mean squared error (MSE) of linear predictors for various data distributions and basis functions as a function of bin width, expressed as a percentage of the total number of sampling points. 

Except for Poisson data with case 2 eigenfunctions, the MSE of linear predictors initially decreases and then increases as bin width grows. A similar trend is observed for the ISE of many eigenfunctions. To further investigate the behavior of Poisson data with case 2 eigenfunctions, we conducted additional experiments at $w = 3\%$ and $4\%$. The resulting MSE values for linear predictors were 0.092 and 0.093, respectively, suggesting that $w = 2\%$ is likely the optimal bin width for this scenario.
\begin{table}[ht]
\begin{center}
\begin{adjustbox}{width=\textwidth}
\begin{tabular}{llcccccccccc}
\toprule
\multicolumn{1}{c}{\,} & \multicolumn{1}{c}{\,} & \multicolumn{5}{c}{Binary} & \multicolumn{5}{c}{Poisson}\\ \cmidrule(lr){3-7}\cmidrule(lr){8-12}
Basis function & Eigenfunction & W = 2\% & W = 5\% & W = 10\% & W = 15\% & W = 20\% & W = 2\% & W = 5\% & W = 10\% & W = 15\% & W = 20\%\\ \midrule
Case 1 & l1e1 & 0.006 & 0.006 & 0.006 & 0.006 & 0.006 & 0.007 & 0.007 & 0.007 & 0.007 & 0.007\\
 & l1e2 & 0.010 & 0.008 & 0.008 & 0.007 & 0.007 & 0.009 & 0.009 & 0.008 & 0.007 & 0.007\\
 & l1e3 & 0.023 & 0.022 & 0.021 & 0.020 & 0.020 & 0.022 & 0.021 & 0.021 & 0.020 & 0.020\\
 & l1e4 & 0.029 & 0.024 & 0.022 & 0.021 & 0.020 & 0.023 & 0.022 & 0.021 & 0.021 & 0.022\\
 & l2e1 & 0.045 & 0.041 & 0.042 & 0.046 & 0.058 & 0.035 & 0.036 & 0.036 & 0.037 & 1.843\\
 & l2e2 & 0.071 & 0.044 & 0.044 & 0.050 & 0.125 & 0.036 & 0.035 & 0.035 & 0.042 & 1.657\\
 & l2e3 & 0.098 & 0.080 & 0.090 & 1.793 & 1.965 & 0.066 & 0.066 & 0.067 & 1.980 & 1.977\\
 & l2e4 & 0.136 & 0.088 & 0.118 & 1.950 & 1.988 & 0.078 & 0.069 & 1.943 & 1.981 & 1.987\\
\midrule Case 2 & l1e1 & 0.004 & 0.004 & 0.006 & 0.014 & 0.023 & 0.003 & 0.003 & 0.005 & 0.011 & 0.020\\
 & l1e2 & 0.011 & 0.009 & 0.008 & 0.007 & 0.008 & 0.006 & 0.005 & 0.005 & 0.005 & 0.006\\
 & l1e3 & 0.010 & 0.010 & 0.018 & 0.046 & 0.083 & 0.010 & 0.011 & 0.018 & 0.044 & 0.075\\
 & l1e4 & 0.039 & 0.015 & 0.011 & 0.014 & 0.027 & 0.009 & 0.007 & 0.007 & 0.012 & 0.025\\
 & l2e1 & 0.111 & 0.014 & 0.005 & 0.002 & 0.001 & 0.006 & 0.002 & 0.001 & 0.001 & 0.001\\
 & l2e2 & 0.090 & 0.015 & 0.006 & 0.004 & 0.005 & 0.003 & 0.002 & 0.002 & 0.003 & 0.004\\
 & l2e3 & 0.166 & 0.027 & 0.015 & 0.022 & 0.035 & 0.010 & 0.007 & 0.009 & 0.019 & 0.031\\
 & l2e4 & 0.175 & 0.034 & 0.038 & 0.080 & 0.123 & 0.014 & 0.016 & 0.030 & 0.071 & 0.113\\
\bottomrule
\end{tabular}
\end{adjustbox}
\end{center}
\caption{ISE of eigenfunctions averaged over 20 repeats for different bin widths.}
\label{supp-table:mise-efs-by-binwidth}
\end{table}

\begin{table}[ht]
\begin{center}
\begin{adjustbox}{width=\textwidth}
\begin{tabular}{lcccccccccc}
\toprule
\multicolumn{1}{c}{\,} & \multicolumn{5}{c}{Binary} & \multicolumn{5}{c}{Poisson}\\ \cmidrule(lr){2-6}\cmidrule(lr){7-11}
Basis function & W = 2\% & W = 5\% & W = 10\% & W = 15\% & W = 20\% & W = 2\% & W = 5\% & W = 10\% & W = 15\% & W = 20\%\\ \midrule
Case 1 & 0.296 & 0.267 & 0.275 & 0.565 & 0.607 & 0.030 & 0.028 & 0.249 & 0.700 & 0.733\\
Case 2 & 0.325 & 0.291 & 0.293 & 0.308 & 0.327 & 0.085 & 0.098 & 0.109 & 0.141 & 0.169\\
\bottomrule
\end{tabular}
\end{adjustbox}
\end{center}
\caption{MSE of linear predictors averaged over 20 repeats for different bin widths.}
\label{supp-table:mise-lpscale-by-binwidth}
\end{table}

\subsection{Effect of the Fixed-effect Intercept}
In this experiment, we focus on binary data and assess the impact of varying the fixed-effect intercept on estimation accuracy while keeping other parameters fixed: sample size $I = 1000$, number of visits $J = 10$, number of sampling points $K = 100$, and percentage of data used for constructing local bins $w = 5\%$. Tables \ref{supp-table:mise-efs-by-intercept} and \ref{supp-table:mise-lpscale-by-intercept} report the integrated squared error (ISE) of eigenfunctions and the mean squared error (MSE) of linear predictors across different values of the intercept. The results indicate that both ISE and MSE improve as the proportion of zeros and ones in the binary data approaches balance (closer to 50\%).
\begin{table}[ht]
\begin{center}
\begin{adjustbox}{width=0.6\textwidth}
\begin{tabular}{llcccc}
\toprule
\multicolumn{1}{c}{\,} & \multicolumn{1}{c}{\,} & \multicolumn{4}{c}{Binary} \\ \cmidrule(lr){3-6}
Basis function & Eigenfunction & b0 =  -3.5 & b0 =  -2.5 & b0 =  -1.5 & b0 =  0\\ \midrule
Case 1 & l1e1 & 0.007 & 0.006 & 0.006 & 0.006\\
& l1e2 & 0.013 & 0.010 & 0.009 & 0.008\\
& l1e3 & 0.028 & 0.024 & 0.023 & 0.022\\
& l1e4 & 0.047 & 0.029 & 0.027 & 0.024\\
& l2e1 & 0.045 & 0.043 & 0.041 & 0.041\\
& l2e2 & 0.076 & 0.054 & 0.047 & 0.044\\
& l2e3 & 0.252 & 0.110 & 0.093 & 0.080\\
& l2e4 & 1.232 & 0.428 & 0.162 & 0.088\\
\midrule Case 2 & l1e1 & 0.006 & 0.005 & 0.005 & 0.004\\
& l1e2 & 0.016 & 0.011 & 0.009 & 0.009\\
& l1e3 & 0.017 & 0.013 & 0.010 & 0.010\\
& l1e4 & 0.674 & 0.107 & 0.024 & 0.015\\
& l2e1 & 0.544 & 0.068 & 0.026 & 0.014\\
& l2e2 & 0.365 & 0.036 & 0.018 & 0.015\\
& l2e3 & 0.500 & 0.086 & 0.040 & 0.027\\
& l2e4 & 0.312 & 0.057 & 0.038 & 0.034\\
\bottomrule
\end{tabular}
\end{adjustbox}
\end{center}
\caption{ISE of eigenfunctions averaged over 20 repeats for different intercepts.}
\label{supp-table:mise-efs-by-intercept}
\end{table}

\begin{table}[ht]
\begin{center}
\begin{adjustbox}{width=0.6\textwidth}
\begin{tabular}{lcccc}
\toprule
\multicolumn{1}{c}{\,} & \multicolumn{4}{c}{Binary} \\ \cmidrule(lr){2-5}
Basis function & b0 =  -3.5 & b0 =  -2.5 & b0 =  -1.5 & b0 =  0\\ \midrule
Case 1 & 0.724 & 0.463 & 0.324 & 0.267\\
Case 2 & 0.895 & 0.545 & 0.378 & 0.291\\
\bottomrule
\end{tabular}
\end{adjustbox}
\end{center}
\caption{MSE of linear predictors averaged over 20 repeats for different intercepts.}
\label{supp-table:mise-lpscale-by-intercept}
\end{table}

\section{More NHANES Data Results}
\subsection{Summary of Demographic Variables and Conventional Mortality Risk Factors}

{\footnotesize
\begin{longtable}{lc} 
\label{supp-table:table1}
% \begin{center}
% \begin{adjustbox}{width=0.5\textwidth}
% \begin{tabular}{lc}
\\ \toprule\toprule
 & Mean (SD) or N (\%)\\\midrule
Sample Size & 4445 \\\midrule
Age   &       64.73 (9.45) \\
Gender & \\
\;\;Female &         2318 (52.1) \\
BMI & \\                                 
\;\;Normal    &  1125 (25.3) \\
\;\;Underweight   &   75 ( 1.7) \\
\;\;Overweight    &  1470 (33.1) \\
\;\;Obese         &  1710 (38.5) \\
\;\;NA            &   65 ( 1.5) \\
Race & \\                                
\;\;Non-Hispanic White   &     1963 (44.2) \\
\;\;Non-Hispanic Black   &     1123 (25.3) \\
\;\;Mexican American     &      428 ( 9.6) \\
\;\;Non-Hispanic Asian   &      427 ( 9.6) \\
\;\;Other Hispanic       &      430 ( 9.7) \\
\;\;Other Race           &       74 ( 1.7) \\
Overall Health           & \\   
\;\;Excellent            &     1743 (39.2) \\
\;\;Very good            &      334 ( 7.5) \\
\;\;Good                 &     1041 (23.4) \\
\;\;Fair                 &      1069 (24.0) \\
\;\;Poor                 &      258 ( 5.8) \\
Diabetes & \\                            
\;\;No                   &     3279 (73.8) \\
\;\;Borderline           &      181 ( 4.1) \\
\;\;Yes                  &      982 (22.1) \\
\;\;NA                   &        3 ( 0.1) \\
Arthritis & \\                           
\;\;No                   &     2459 (55.3) \\
\;\;Yes                  &     1975 (44.4) \\
\;\;NA                   &       11 ( 0.2) \\
CHF & \\                                 
\;\;No                   &     4162 (93.6) \\
\;\;Yes                  &      268 ( 6.0) \\
\;\;NA                   &       15 ( 0.3) \\
CHD & \\                                 
\;\;No                   &     4076 (91.7) \\
\;\;Yes                  &      341 ( 7.7) \\
\;\;NA                   &       28 ( 0.6) \\
Heart Attack & \\                        
\;\;No                   &     4107 (92.4) \\
\;\;Yes                  &      332 ( 7.5) \\
\;\;NA                   &        6 ( 0.1) \\
Stroke & \\                              
\;\;No                   &     4121 (92.7) \\
\;\;Yes                  &      318 ( 7.2) \\
\;\;NA                   &        6 ( 0.1) \\
Cancer & \\                              
\;\;No                   &     3707 (83.4) \\
\;\;Yes                  &      737 (16.6) \\
\;\;NA                   &        1 ( 0.0) \\
Alcohol Consumption & \\             
\;\;Never drinker       &       697 (15.7) \\
\;\;Former drinker      &      1081 (24.3) \\
\;\;Moderate drinker    &      2161 (48.6) \\
\;\;Heavy drinker       &       257 ( 5.8) \\
\;\;Not Available       &       249 ( 5.6) \\
Smoking & \\                   
\;\;Never               &      2224 (50.0) \\
\;\;Former              &      1477 (33.2) \\
\;\;Current             &       741 (16.7) \\
\;\;NA                  &         3 ( 0.1) \\
Mobility Problem & \\                    
\;\;No difficulty       &      2929 (65.9) \\
\;\;Any difficulty      &      1511 (34.0) \\
\;\;NA                  &         5 ( 0.1) \\
\bottomrule\bottomrule
% \end{tabular}
% \end{adjustbox}
% \end{center}
\\ \caption{Summary of demographic variables and traditional risk factors for subjects over the age of 50 from the NHANES 2011-2014 accelerometry data set.}
\end{longtable}}

\subsection{Choosing Bin Width}
\begin{figure}[!tbh]
\centering
  \includegraphics[width=0.75\textwidth]{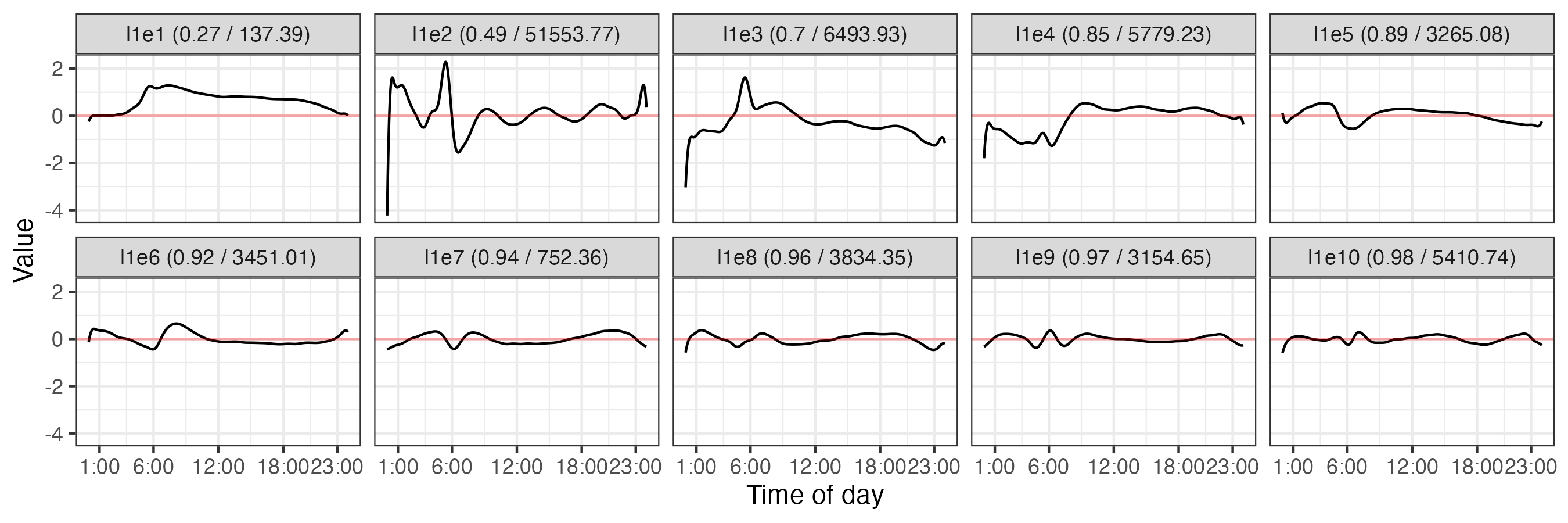}
\includegraphics[width=0.75\textwidth]{figures/nhanes-2011/nhanes-2011-l1ef-binwidth30.png}
  \includegraphics[width=0.75\textwidth]{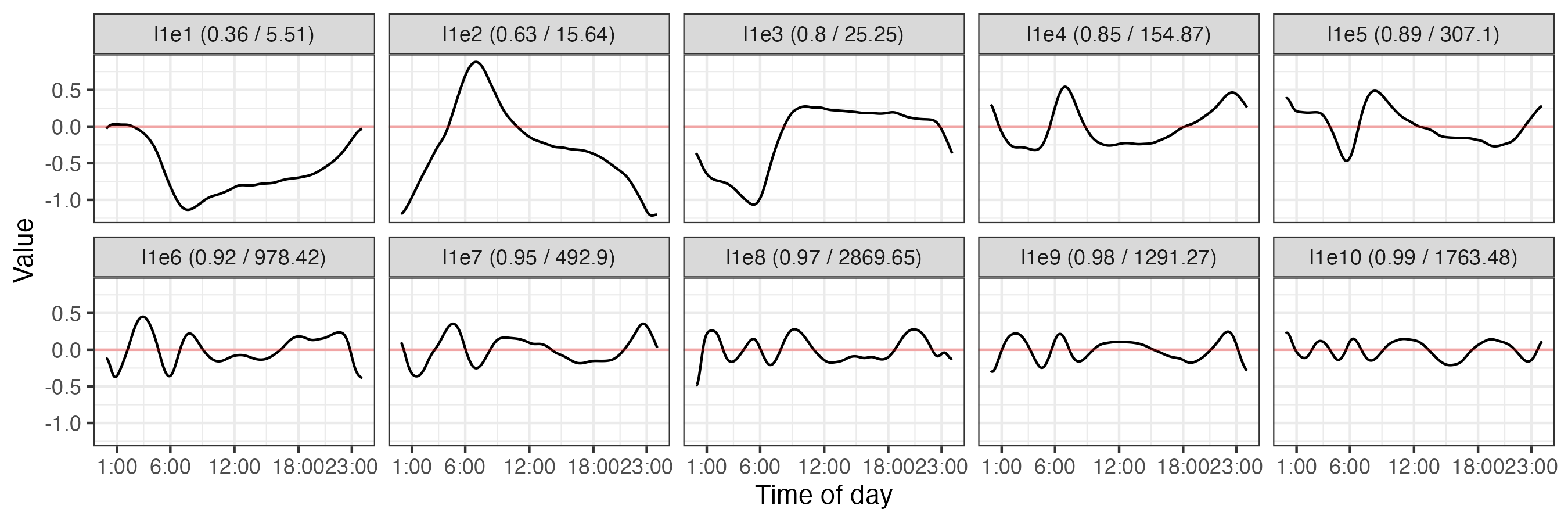}
  \caption{Level 1 eigenfunctions obtained under bin widths 10 minutes (top), 30 minutes (middle), and 60 minutes (bottom) for the NHANES data.}
  \label{supp-fig:nhanes-l1ef-diff-binwidth}
\end{figure}

\begin{figure}[!tbh]
\centering
  \includegraphics[width=0.75\textwidth]{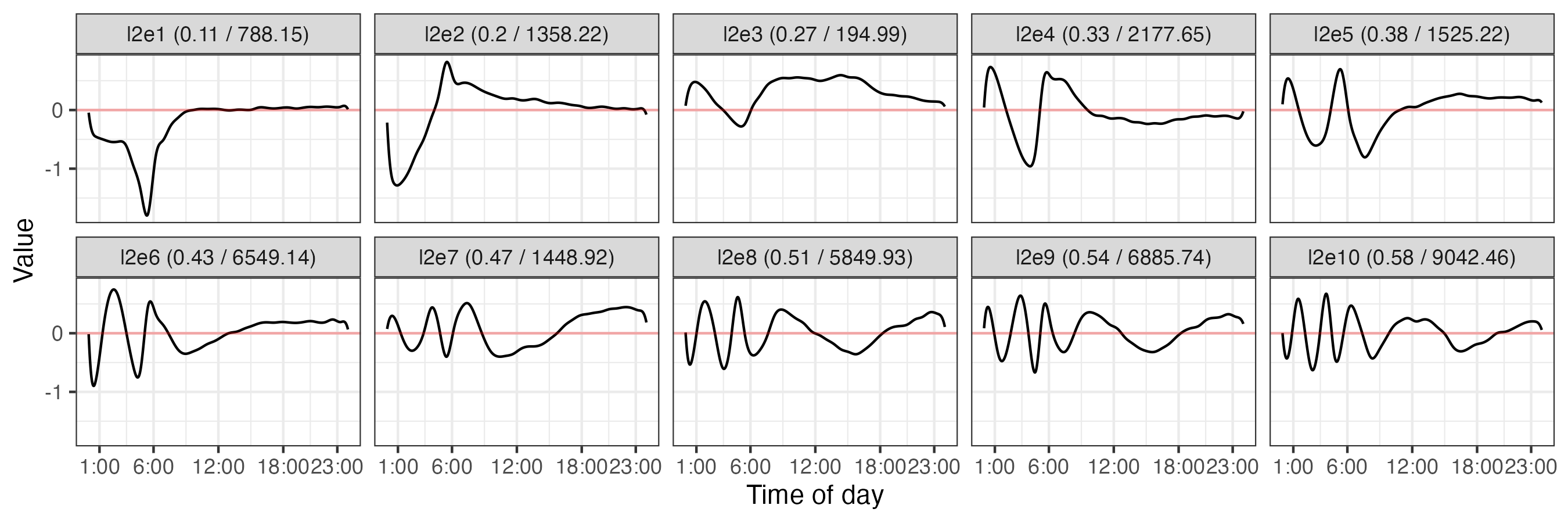}
\includegraphics[width=0.75\textwidth]{figures/nhanes-2011/nhanes-2011-l2ef-binwidth30.png} 
  \includegraphics[width=0.75\textwidth]{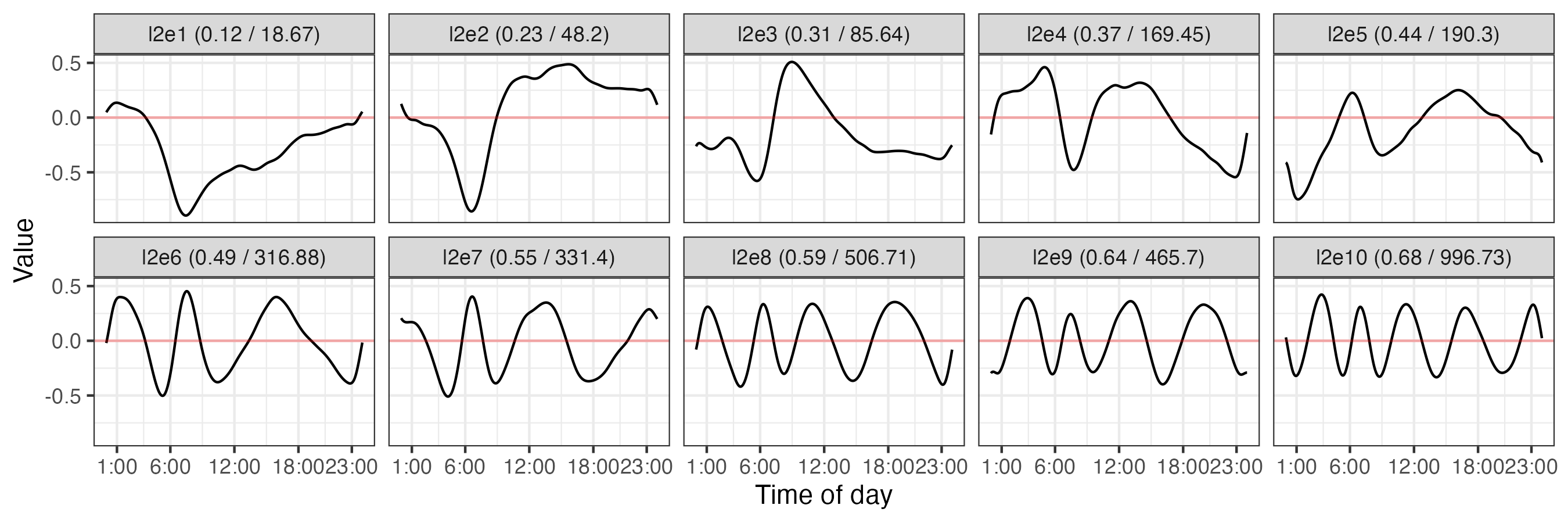}
  \caption{Level 2 eigenfunctions obtained under bin widths 10 minutes (top), 30 minutes (middle), and 60 minutes (bottom) for the NHANES data.}
  \label{supp-fig:nhanes-l2ef-diff-binwidth}
\end{figure}

To determine an appropriate bin width for the NHANES data, we examined three values: 10 minutes, 30 minutes, and 60 minutes. We fit local GLMMs using each bin width and compared the estimated eigenfunctions. Figures~\ref{supp-fig:nhanes-l1ef-diff-binwidth} and \ref{supp-fig:nhanes-l2ef-diff-binwidth} display level 1 and level 2 eigenfunctions, respectively. We observe that results using 30-minute and 60-minute bin widths are similar. However, with a 10-minute bin width, some eigenfunctions (e.g., l1e1, l1e5) exhibit similarities to those estimated with larger bin widths (possibly differing in sign), whereas others deviate significantly. Moreover, when using a 10-minute bin width, we encountered singular fits in 31 bins, likely due to numerous subject-visit pairs having all-zero in those bins during nighttime. Increasing the bin width to 30 minutes eliminated singular fits.

\newpage

\bibliographystyle{apalike} % Style BST file
\bibliography{ref}       % Bibliography file (usually '*.bib')
\end{document}